\documentclass[%
 reprint,
 amsmath,amssymb,
 aps,
]{revtex4-2}
\usepackage{lipsum}
\usepackage{comment}
\usepackage{graphicx}% Include figure files
\usepackage{xcolor}%
\usepackage{dcolumn}% Align table columns on decimal point
\usepackage{bm}% bold math
\usepackage{array} % in the preamble
\usepackage{subfigure}
\usepackage{hyperref}
\usepackage{url}
\usepackage{hyperref}
\usepackage{adjustbox}

\usepackage{hyperref}

\begin{document}
%\begin{comment}

\preprint{APS/123-QED}

\title{Quantum Generative Learning for High-Resolution Medical Image Generation}% Force line breaks with \\

\author{Amena Khatun\textsuperscript{1}}
\email{amena.khatun@data61.csiro.au}
\author{K\"ubra Yeter Aydeniz\textsuperscript{2}}
\author{ Yaakov S. Weinstein\textsuperscript{2}}
\author{Muhammad Usman\textsuperscript{3, 4}}

\affiliation{%
\textsuperscript{1}Data61, CSIRO, Dutton Park, QLD 4102, Australia\\
\textsuperscript{2}Emerging Engineering and Physical Sciences Department, The MITRE Corporation, USA\\
\textsuperscript{3}Data61, CSIRO, Research Way, Clayton, 3168, Victoria, Australia\\
\textsuperscript{4}School of Physics, The University of Melbourne, Parkville, 3010, Victoria, Australia
}%
 
%\author{Muhammad Usman}
%\email{muhammad.usman@data61.csiro.au}
%\affiliation{%
%%%%%School of Physics, The University of Melbourne, Parkville, 3010, Victoria, Australia
%%%%}%
%%%\affiliation{%
%%Data61, CSIRO, Research Way, Clayton, 3168, Victoria, Australia
%}%

\begin{abstract}
Integration of quantum computing in generative machine learning models has the potential to offer benefits such as training speed-up and superior feature extraction. However, the existing quantum generative adversarial networks (QGANs) fail to generate high-quality images due to their patch-based, pixel-wise learning approaches. These methods capture only local details, ignoring the global structure and semantic information of images. In this work, we address these challenges by proposing a quantum image generative learning (QIGL) approach for high-quality medical image generation. Our proposed quantum generator leverages variational quantum circuit approach addressing scalability issues by extracting principal components from the images instead of dividing them into patches. Additionally, we integrate the Wasserstein distance within the QIGL framework to generate a diverse set of medical samples. Through a systematic set of simulations on X-ray images from knee osteoarthritis and medical MNIST datasets, our model demonstrates superior performance, achieving the lowest Fréchet Inception Distance (FID) scores compared to its classical counterpart and advanced QGAN models reported in the literature.

\end{abstract}

\maketitle

%\tableofcontents

\begin{figure*}
\begin{center}
\includegraphics[width=1.0\linewidth]{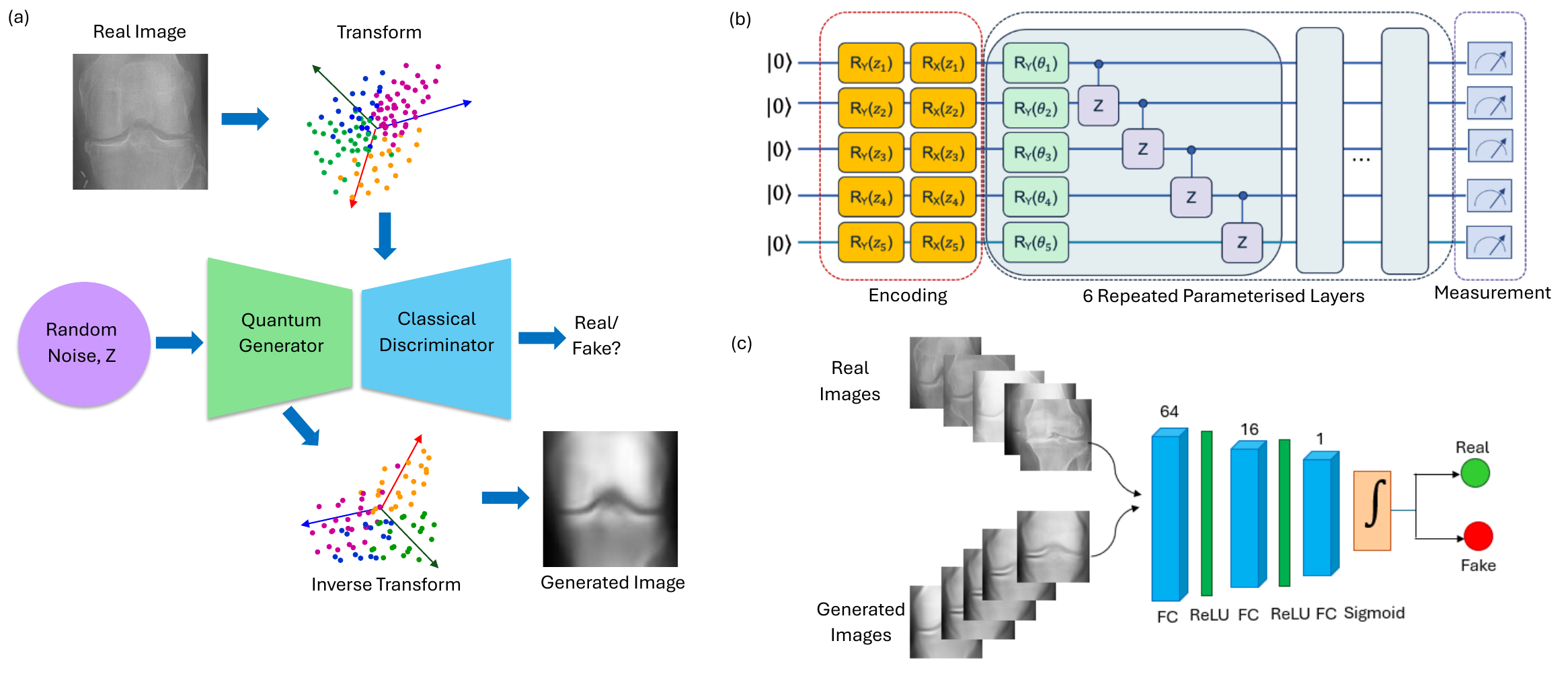}
\end{center}
\caption{(a) Overview of Quantum Image Generative Learning (QIGL) framework. QIGL consists of two sub-networks: a generator and a discriminator. Here, the generator is quantum and the discriminator is classical. The real input images are transformed before being fed into the discriminator. The quantum generator initiates the learning process to replicate the distribution throughout the training of QIGL. The PCA features are then re-scaled to revert to the original principal component range and produce a distinctive image that aligns with the distribution of the original images. (b) Architecture of quantum generator that is composed of eight sub-generators, each implemented as a variational quantum circuit (VQC). Each VQC consists of fixed gates for entangling the qubits and tunable gates optimized through training. The generator circuit's architecture begins with encoding input noise using $R_X(\theta_i)$ and $R_Y(\theta_i)$ rotational gates. Subsequently, parameterized weights are encoded on each quantum layer, accompanied by $CZ$ gates to entangle qubits within each layer. Following these repeated layers, the expected value of Pauli $X$ is computed for each qubit upon measurement. (c) Classical discriminator initiates with a dense layer (fully connected (FC)) for linear transformation of the input data that projects the input tensor into 64 dimensions. Subsequently, a Rectified Linear Unit (ReLU) activation function is applied element-wise to the output tensor. Following this, the output of the final dense layer passes through a Sigmoid activation function for final prediction whether the data is real or fake.}
\label{fig:QGAN_overview}
\end{figure*}
\section{\label{sec:level1}Introduction}

Medical images are commonly used in diagnostics, neuroscience and drug discovery, with the use of artificial intelligence (AI) in these applications becoming increasingly common \cite{esteva2017dermatologist, litjens2017survey, stokes2020deep}. Despite the promise of AI to improve efficiency and accuracy, the challenge is in its training due to limited availability of medical images and patient datasets. A potential solution is, therefore, the generation of synthetic data using generative adversarial networks (GANs). To generate synthetic images which have similar properties to the original images from patient scans, GANs were first introduced in Ref. \cite{goodfellow2014generative}. Recently, GANs have demonstrated state-of-the-art \cite{khatun2021end, khatun2023pose} capabilities in generating realistic, high-resolution image datasets \cite{karras2020analyzing, brock2018large}, among other purposes including text generation \cite{iqbal2022survey}, data augmentation \cite{Radford2015UnsupervisedRL}, and anomaly detection \cite{pang2021deep}. Despite their widespread success, GANs impose huge computational demands, especially to generate high-resolution images \cite{, brock2018large, karras2019style}. Contrarily, the integration of quantum computing in GAN architecture design could overcome these computational challenges. Indeed the design and implementation of resource efficient quantum generative adversarial networks (QGANs) is an emerging and promising research direction.

Existing work on QGANs \cite{lloyd2018quantum, dallaire2018quantum} have demonstrated that QGANs can achieve convergence similarly to classical GANs but within a quantum framework. These studies also highlight the advantages of QGANs in efficiently processing high-dimensional data, making them powerful for certain applications. However, these studies are often limited to theoretical proofs \cite{lloyd2018quantum, dallaire2018quantum}, inefficient data loading \cite{nakaji2021quantum}, dealing with small-scale datasets \cite{huang2021experimental}, exhibiting instability during training \cite{huang2021experimental}, and suffering from mode collapse \cite{huang2021experimental}, generally leading to poor-quality image generation. The advanced QGANs for image generation were proposed in Ref. \cite{huang2021experimental}. This study considered a handwritten digits dataset ($8\times8$ pixels) for image generation of digits $0$ and $1$. Although it could generate recognisable images in some cases, the quality  the images was generally poor. As this approach divided images into patches and performed pixel-wise learning, it suffered from scalability challenges, particularly for high-resolution images. \textcolor{black}{This leads to the critical question: Can a QGAN effectively generate high-dimensional, diverse, and high-quality samples that are scalable to real-world applications?}

\textcolor{black}{To overcome the limitations of existing QGAN methods, this work proposes a quantum image generative learning (QIGL) to generate high-quality images while overcoming scalability issues and mitigating mode collapse.} We implement and benchmark our QIGL technique on medical image datasets which typically require good quality image generation for diagnostic purpose. Figure \ref{fig:QGAN_overview}(a) illustrates QIGL approach that consists of two sub-networks as in traditional GAN framework. The generator is quantum and the discriminator is classical. This is similar concept to other QGAN implementations \cite{huang2021experimental, tsang2023hybrid}. The architectures of the generator and discriminator networks are illustrated in Figures \ref{fig:QGAN_overview}(b) and \ref{fig:QGAN_overview}(c). The variational quantum circuit-based generator is trained on quantum resources to exploit the unique properties of quantum mechanics during both the learning process and the generation of new samples. The discriminator is trained on classical resources that distinguishes the real and fake samples and is not required once the model is trained. The real input images undergo a principal component analysis (PCA) transformation before being processed by the discriminator. The quantum generator spearheads the learning process, aiming to reproduce the distribution of the original image dataset. After generating the images, both the original and the generated samples are fed to the discriminator. This adversarial setup ensures that the generator and discriminator continually learn from each other, improving their capabilities until the generator produces high-fidelity images indistinguishable from real data. Our work shows that QIGL generates high-quality medical images while mitigating scalability and mode collapse issues that are present in the current methods. 

The generation of realistic medical images also contributes to the development of algorithms used in medical image analysis and diagnosis. To this end, we consider knee osteoarthritis (OA) dataset \cite{chen2018knee} that consists of 5 classes, where the original image dimension is $224\times224$. The class labels of the knee OA dataset were identified based on the Kellgren-Lawrence grading system (KL-grade) that is used to define knee OA progression, to assess OA in epidemiologic studies and to provide documentation for the requirement of total knee replacement \cite{9098456}. In addition to this, we studied the medical MNIST dataset, containing 6 classes with image dimensions of 64$\times$64. QIGL learns and extracts rich principal components from the images, resulting in the generation of high-quality and high-dimensional medical images. Additionally, we address the mode collapse issue by incorporating the Wasserstein distance \cite{arjovsky2017wasserstein} within the proposed framework as a loss function to enable better convergence and generating diverse and realistic samples.

\section{\label{sec:level2}Related Work}
In 2014, Goodfellow introduced generative models in an adversarial manner \cite{goodfellow2014generative}, where generator captures the data distribution, and a discriminator model estimates the probability that a sample came from the training data rather than generative model. While it is challenging,  having a model to capture the distribution of images to generate realistic synthetic copies, GANs achieved great success synthetic for image generation. Inspired by the success of GANs, various extensions have been proposed for image-to-image translation \cite{khatun2020semantic, gatys2016image}, pixel-level transfer from source to target domains \cite{yoo2016pixel}, and style transfer between domains \cite{khatun2023pose}.

Witnessing the success of classical GANs, quantum information science community has been motivated to develop corresponding quantum GANs that can run on quantum computer. Ref. \cite{lloyd2018quantum} first introduced QGANs, demonstrating their theoretical potential for exponential superiority compared to its classical counterparts. Building on this, Ref. \cite{dallaire2018quantum} designed a way to train variational quantum circuit (VQC) as the generator in QGAN models. Although these methods are only proof-of-concept,  the findings of these studies highlight the potential of QGANs to address problems that classical GANs struggle with includes huge demand of computational resources and the number of parameters, which increase the complexity of the model, making optimisation and convergence more challenging \cite{heusel2017gans}. 

Ref. \cite{qu2023quantum} introduces a patch-based conditional QGAN for generating abnormal ECG signals. The first image generation QGAN method was proposed in Ref. \cite{huang2021experimental} that generate images of hand-written digits from MNIST dataset. This approach down-scaled the size of the images from $28 \times 28$ into $8 \times 8$ and divided the images into patches before using as input to the generator as lower dimensional images requires less number of qubits. Despite significant performance gain in synthetic image generation with fewer parameters compared to classical GANs, this method suffers from scalability issues, especially for high-resolution images as it fails to capture the global patterns and semantic details of the images due to pixel-wise learning. 1) By concentrating on individual patches and pixels, the model may lose sight of the broader context and relationships between different parts of the image. This leads to difficulties in capturing global patterns and semantic details. 2) Processing images pixel by pixel can be computationally expensive and memory-intensive, particularly for high-resolution images. 3) In many real-world applications, it is essential to consider the entire image and understand the relationship between objects and features across the image. Pixel-wise learning may not effectively capture these broader relationships. 4) Pixel-wise learning is not effective at generating a diverse set of images as it only consider the similarity between individual pixels, while ignoring high-level features and semantics of the images. This leads to a loss of diversity in the generated samples, which is known as mode collapse problem. Another study \cite{tsang2023hybrid} uses a similar concept as in Ref. \cite{huang2021experimental} and generates images with $28 \times 28$ pixel rather than reducing the dimensionality as in \cite{huang2021experimental}. However, this method also breaks the images into patches and performs pixel-wise learning, which leads to the scalability issues and poor quality image generation. Ref. \cite{zhou2023hybrid} introduced QGAN via learning discrete distribution for image generation. A remapping method is also used to enhance the quality of the generated images of MNIST and Fashion-MNIST datasets, however, this approach also generate low resolution images (28 $\times $ 28 pixel). Ref. \cite{silver2023mosaiq} also proposed a QGAN method to generate low-resolution images for MNIST and Fashion-MNIST datasets. {\color{black}{While the existing QGANs focus on low resolution image generation tasks, Refs.  \cite{landman2022quantum} and \cite{cherrat2024quantum} explore medical image processing using quantum neural networks for medical image classification.}} \textcolor{black}{Ref. \cite{stein2021qugan} involves dimensionality reduction from 784 to 4, however, the core idea of this approach is a swap test for measuring the quantum state difference, which is based on quantum-state loss functions and operates through a single ancilla qubit. The method is then evaluated on MNIST dataset for over 99 epochs for image generation. This approach fundamentally differs from QIGL, which is built upon a hybrid quantum-classical framework with a standard classical GAN concept.  In QIGL, we use multiple quantum sub-generators (8 in total) that distribute the 40 PCA features to ensure the preservation of high-level information across the image generation process. The feature redistribution further allows for better scalability and quality of generated images, particularly for high-resolution medical datasets, while the Wasserstein distance enables better convergence and diversity in generated images. To further enhance the quality of the images, we perform histogram equalisation (HE) that spreads out the most frequent intensity values in an image, making it easier
for the model to learn and generate more detailed and
visually appealing images. It also helps in reducing the
model’s bias towards certain intensity values, thereby enhancing
the overall quality of the generated images.}

\textcolor{black}{Despite recent progress in developing QGANs for image generation, their scope is still limited due to low resolution and proof-of-concept \cite{lloyd2018quantum, dallaire2018quantum} dataset and  scalability issues \cite{huang2021experimental}, especially
for high-resolution images as it fails to capture the
global patterns and semantic details of the images due
to pixel-wise learning. Hence, the scope of these methods are limited and impractical for real-world applications. Our work addresses this critical gap  by extracting and learning the essential features of the input images rather constructing the pixels of an image directly and consequently. We ensure to capture the most significant variations (98\%) within the dataset by extracting 40 PCA features which captures 98\% of the total variability present in the dataset, and hence, preserving the high-level features across the images. Through systematic experimental simulations, we demonstrates that quantum generative learning could generate high resolution images and could have a direct application in the field of medical diagnostic, which is an important application with real-world implications.}

\section{Methods}
\label{Proposed Method}
In this section, we introduce QIGL, detailing its design and architecture. First, we briefly describe classical GAN for better context of understanding. Subsequently, we delve into the details of QIGL.

GANs work through a dynamic interplay between two neural networks: the generator and the discriminator. The generator, creates synthetic data, while the discriminator tries to distinguish between real and fake samples. As they compete, both networks improve iteratively. The generator aims to produce samples that are indistinguishable from real data, while the discriminator aims to become increasingly accurate at differentiating between real and fake samples. Through this adversarial process, GANs can generate highly realistic data. This is explained in the supplementary document in detail (see \ref{Sup-classicalGAN}).

\subsection{Quantum Image Generative Learning (QIGL)}
The proposed QIGL consists of a generator and a discriminator, following a similar structure to classical GANs, however, utilising quantum principles for training and inference. The generator is trained on quantum resources while the discriminator operates on classical resources, allowing QIGL to leverage quantum resources for generative tasks. To address the limitations of current NISQ (Noisy Intermediate-Scale Quantum) devices, which have fewer qubits than required for high-resolution image generation, we employ a PCA-based approach. PCA extracts essential features of images, which are scaled between 0 and 1, enabling quantum sub-generators to operate efficiently. The scaled features guide the generator during training. Once training is completed, the output is re-scaled to its original range and inverse-transformed to reconstruct high-quality images. To ensure balanced feature distribution among sub-generators, the top principal components are assigned evenly, avoiding the imbalance and enhancing the overall utility of the generator. This methodology overcomes scalability issues and generates high-quality medical images with minimal qubits.

\subsection{Quantum Generator and Classical Discriminator Framework}
\label{generator_discriminator}
In this work, we propose QIGL composed of a quantum generator and a classical discriminator as illustrated in Figure \ref{fig:QGAN_overview}(a), where the quantum generator consists of multiple sub-generators. Each of these sub-generators is a variational quantum circuit (VQC), which undergoes iterative optimisation to train the model. VQC consists of fixed gates for entangling the qubits and the tunable gates that are optimised through the training process. Thus, the complete quantum circuit is an optimisation function constructed from a series of unitary transformations. The architecture of a generator circuit is illustrated in Figure \ref{fig:QGAN_overview}(b). The generator takes random noise vectors as input, where each element is a random value. \textcolor{black}{The elements of noise vector $Z$  are sampled from a uniform distribution over the interval $[0, \pi/2]$. This choice ensures an appropriate range for the quantum rotational gates $(R_X, R_Y)$, which are parameterised by these noise values. The uniform distribution is chosen because it provides a balanced range of values that can effectively explore the parameter space of the quantum circuit, facilitating the learning process.
Mathematically, each element $z_i$ in $Z$ is drawn as:
\begin{align}
z_i \sim \text{Uniform}(0, \pi/2).
\end{align}
This noise vector is then used to initialize the qubits in the quantum circuit via \( R_X \) and \( R_Y \) gates:
\begin{align}
R_X(z_i) = e^{-i z_i X / 2}, \quad R_Y(z_i) = e^{-i z_i Y / 2},
\end{align}
where X and Y are the Pauli-X and Pauli-Y matrices, respectively.} Each element of the noise vector corresponds to a qubit in the quantum circuit. Each qubit is initialised with $R_x$ and $R_y$ rotational gates, parameterised by the noise input. Subsequently, after the noise has been embedded, parameterised weights are encoded on each quantum layer alongside controlled-$Z$ (CZ) gates to entangle the qubits within each layer. After these six repeated layers, the $Pauli-X$ expected value is computed for each qubit upon measurement. For PCA feature selection, we choose 40 components as these 40 components capture 98\% of the total variability present in the dataset (see Section \ref{conventionalPCA} and Figure \ref{fig:scree-plot} in the supplementary document). These 40 principal components are distributed across  eight sub-generators. \textcolor{black}{All the eight sub-generators have the similar structures as described above, each comprising a five-qubit circuit. Each sub-generator operates independently and is parameterised by trainable weights, enabling it to learn distinct aspects of the data distribution. The use of multiple sub-generators allows the model to parallelise computation and scale efficiently within the limitations of current quantum hardware. Each sub-generator consists of random noise which is encoded into the quantum state using the rotational gates, a sequence of parameterised rotations and $CZ$ entangling gates across 6 layers (circuit depth), and the expectation value of the Pauli-X operator is measured for each qubit. Each sub-generator processes its assigned five PCA features and outputs five expectation values from its five qubits. The total output from all sub-generators is a concatenated vector of $8\times5=40$ values, matching the 40 PCA components. These 40 features are reordered to match the original PCA component hierarchy. This step ensures that the generated features align with the variance structure of the training data. During training, the parameters of all sub-generators are optimised jointly through backpropagation, guided by the Wasserstein loss function. This ensures that the sub-generators collaborate to reproduce the global distribution of the original dataset.}

Figure \ref{fig:QGAN_overview}(b) shows the structure of only one sub-generator. \textcolor{black}{Each sub-generator outputting Pauli-X expectation values corresponding to a subset of PCA-transformed image features. These expectation values represent the quantum state of the qubits after the circuit has been executed. The expectation values from all sub-generators are post-processed to form the output of the quantum generator. Specifically, the expectation values are flattened and rearranged according to the predefined ordering of the PCA components. This ensures that the generated data aligns with the structure of the original dataset. Once the expectation values are post-processed, they are passed through an inverse PCA transformation to reconstruct the generated images. The inverse PCA transformation maps the reduced-dimensional data (40 principal components) back to the original image space ($64\times64$ pixels). Mathematically, let $m=[m_1, m_2,....,m_{40}]$ be the concatenated Pauli-X expectation values from all eight sub-generators, where each sub-generator produces 5 expectation values. The final reconstructed image is obtained via the inverse PCA transformation:
\begin{align}
x_{\text{gen}} = U_k \Sigma_k m + \mu,
\end{align}
where $U_k$ and $\Sigma_k$ are the matrices containing the first $k$ principal components and their corresponding singular values, respectively. $\mu$ is the mean of the original data used during PCA training. This process ensures that the generated images retain the global structure and semantic details of the original dataset, as the PCA transformation captures the most significant variations in the data. }

\textcolor{black}{The pixel values of the original images are flattened into a 1D vector, where each image is represented as a vector of length 4096 (for $64\times64$ pixels). PCA is applied to the preprocessed image data to extract the principal components. The number of principal components is chosen such that they capture a significant portion of the total variance (in this case, 98\% of the variance is captured by 40 principal components). The mathematical explanation have been added in the supplementary document. The PCA-transformed data is incorporated into the loss function through the training process of the QIGL framework. The quantum generator takes a random noise vector as input and produces a set of expectation values. These expectation values are post-processed and reshaped to match the dimensions of the PCA-transformed data (40 features).} \textcolor{black}{The loss function used in the QIGL framework is the Wasserstein distance, which measures the dissimilarity between the distributions of real and fake data. The loss function is constructed using the  Kantorovich-Rubinstein duality as,
\begin{align}
W(P_r, P_{\theta}) = \sup_{\|f\|_{L} \leq 1} \mathbb{E}_{x \sim P_r} [f(x)] - \mathbb{E}_{x \sim P_{g}} [f(m)],
\end{align}
where $P_r$ represents the real data distribution in PCA space, $P_g$ represents the generated data distribution in PCA space which is Pauli-X expectation values from the quantum generator. $f(x)$ is the discriminator which learns to approximate the optimal 1-Lipschitz function in the Kantorovich-Rubinstein duality. The supremum (sup) is taken over all functions $f$ with $\|f\|_{L} \leq 1$. $\mathbb{E}_{x \sim P_r} [f(x)]$ represents the expected value of $f(x)$) when $x$ is sampled from the real data distribution $P_r$, and $\mathbb{E}_{x \sim P_g} [f(m)]$ represents the expected value of $f(x)$ when $x$ is sampled from the generated data distribution, $P_g$, which is controlled by the generator network. We approximate this distance by training a discriminator to distinguish between real and generated distributions. The discriminator receives two types of inputs: the PCA-transformed real images are used as the ground truth and the output of the quantum generator, which is also in the PCA-transformed space, is used as the fake data. The loss function for the discriminator is given by:
\begin{align}
L_D = - \mathbb{E}_{x \sim P_r} [D(x)] + \mathbb{E}_{m \sim P_g} [D(m)],
\end{align}
$D(\cdot)$ is the discriminator function that differentiates real PCA features from generated features.}
\textcolor{black}{The loss function for the generator is given by:
\begin{align}
L_G = - \mathbb{E}_{m \sim P_g} [D(m)]
\end{align}
The quantum generator learns to output Pauli-X expectation values $m$ that approximate the PCA feature distribution of real images. The discriminator and generator are optimized iteratively to minimise the Wasserstein distance, ensuring that the generated data matches the distribution of real data in the PCA-transformed space.}

The classical discriminator architecture is illustrated in Figure \ref{fig:QGAN_overview}(c), where the first layer is a dense layer that performs linear transformation of the input data. It takes the input tensor with 40 PCA dimensions and projects it into a space with 64 dimensions. Each neuron in this layer is connected to every neuron in the input, applying weights to the input features to transform them into a new representation. After that, a Rectified Linear Unit (ReLU) activation function is applied element-wise to the output tensor. ReLU introduces non-linearity by replacing negative values with zero which helps the network learn complex patterns and relationships in the data. The output of the final dense layer is passed through a Sigmoid activation function to transform the output of the network into a probability value ranges between $0$ and $1$. Values closer to 1 indicate higher probability of the input being classified as real, whereas values closer to 0 indicate the probability of the input being classified as generated. \textcolor{black}{ The discriminator must satisfy a 1-Lipschitz condition to ensure the Wasserstein distance is valid. We impose this constraint using weight clipping, where all the weights of the discriminator are constrained within a fixed range. This prevents gradient explosion and ensures stable training.}

\textcolor{black}{
The total objective function is the min-max optimisation problem:
\begin{align}
\min_G \max_{D \in \mathcal{D}} \left( - \mathbb{E}_{x \sim P_r} [D(x)] + \mathbb{E}_{m \sim P_g} [D(m)] \right).
\end{align}
}

\subsection{Network Setup and Training Details}
We use PyTorch \cite{paszke2019pytorch} as a wrapper to interface with PennyLane \cite{bergholm2018pennylane}. Adam \cite{kingma2014adam} optimiser is used for all the experiments with a batch size of 8 and a learning rate of 0.3 for the quantum generator and 0.05 for the discriminator. We initialise the quantum circuit with 5 qubits and 6 variational layers (depth of the quantum circuit). During training, images are decomposed into 40 principal components which are then distributed across eight 5-qubit quantum circuit, each of which is a sub-generator. We also perform simulation studies for Quantum PatchGAN \cite{huang2021experimental} and PQWGAN \cite{tsang2023hybrid} on Knee OA dataset for comparison. For these implementations, we use Pennylane framework with the same hyper-parameters used in the original paper, i.e., Ref.~\cite{huang2021experimental, tsang2023hybrid}. To further evaluate the proposed method, we additionally set up simulations for classical WGAN with the same hyper-parameters as used in QIGL implementation for fair evaluation. \textcolor{black}{The QIGL model was trained for 50 epochs on a high-performance computing (HPC) system with a single NVIDIA GPU. The training time varies depending on the dataset size; however, for the Knee OA dataset, the total training duration was approximately 3 hours, 8 minutes, and 34 seconds. On average, each epoch took around 210.53 seconds (approximately 3.5 minutes) to complete.}

\subsection{Image Pre-processing}
\label{image_processiong}
In order to improve the quality of the generated images we implemented image preprocessing. One of the methods used in image generation through GANs is to use histogram equalisation which improves the contrast of the generated images. This technique spreads out the most frequent intensity values in an image, making it easier for the model to learn and generate more detailed and visually appealing images. It also helps in reducing the model's bias towards certain intensity values, thereby enhancing the overall quality of the generated images. In order to implement this technique, we utilized \textit{equalize} function in Python Imaging Library's (PIL) \textit{ImageOps} module. This function applies a non-linear mapping to the input image in order to create a uniform distribution of grayscale values in the output image. \textcolor{black}{Histogram equalization enhances contrast by redistributing pixel intensities according to the cumulative distribution function (CDF) of the image histogram. his is particularly beneficial for medical images (e.g., X-rays), where subtle intensity variations may correspond to critical diagnostic features. For an image with intensity values $i \in [0, 255]$, the HE transformation is defined as:
\begin{align}
T(i) = \text{round} \left( 255 \times \sum_{j=0}^{N} h(j) \right),
\end{align}
where $h(j)$ is the probability of the intensity values of the images, $N$ is the total number of pixels. This pre-processing step ensures that the discriminator receives images with enhanced detail and reduced intensity bias, improving the generator’s ability to learn realistic features.} Another image preprocessing method that we implemented is to manually detect and remove the images with surgical prosthetics or scratches and punch holes. In addition to this, we inverted negative channel images and flipped all right-oriented images to the left, as it was implemented in \cite{prezja2022deepfake}. In Table \ref{tab:preprocessing} in the supplementary document, we provide the number of these items in each KL grade of osteoarthritis of the knee X-ray images dataset.

\section{Results and Discussion}
QIGL is implemented and tested on two popular medical image datasets: knee OA, consists of 5 grade/classes (healthy, doubtful, minimal, moderate, and severe), and medical MNIST dataset, comprising 6 classes (abdomen CT, breast MRI, chest CT, chest X-ray, hand, and head) which are shown in Fig. \ref{fig:Datasets} in the supplementary document and briefly discussed in this section.
\label{Results}

\subsection{Datasets}
\label{sec:datasets}
\textbf{Knee Osteoarthritis} \cite{chen2018knee} dataset includes knee X-ray image data aimed at both knee joint detection and grading according to the Kellgren-Lawrence (KL) scale,  facilitating research and analysis in the diagnosis of knee osteoarthritis. For image generation, we consider the grading as different class. There are 5 grading system within this dataset such as: Grade 0 represents healthy knee images, Grade 1 indicates doubtful joint narrowing with possible osteophytic lipping, Grade 2 reflects definite presence of osteophytes with potential joint space narrowing, Grade 3 signifies multiple osteophytes, definite joint space narrowing, and mild sclerosis, and Grade 4 denotes large osteophytes, significant joint narrowing, and severe sclerosis. Total 5,778 images in 224$\times$224 dimension are considered from all 5 classes to train QIGL model for image generation.

\textbf{Medical MNIST} \cite{MedicalMNISTClassification} dataset is a MNIST-style medical images in 64$\times$64 dimension. The images were originally taken from other datasets and processed into such style. There are 58954 medical images belonging to 6 classes, i.e., AbdomenCT, BreastMRI, ChestCT, ChestXR, Hand, and HeadCT. To train QIGL, all these images are considered as real input images.

\begin{figure}
\begin{center}
\includegraphics[width=1.0\linewidth]{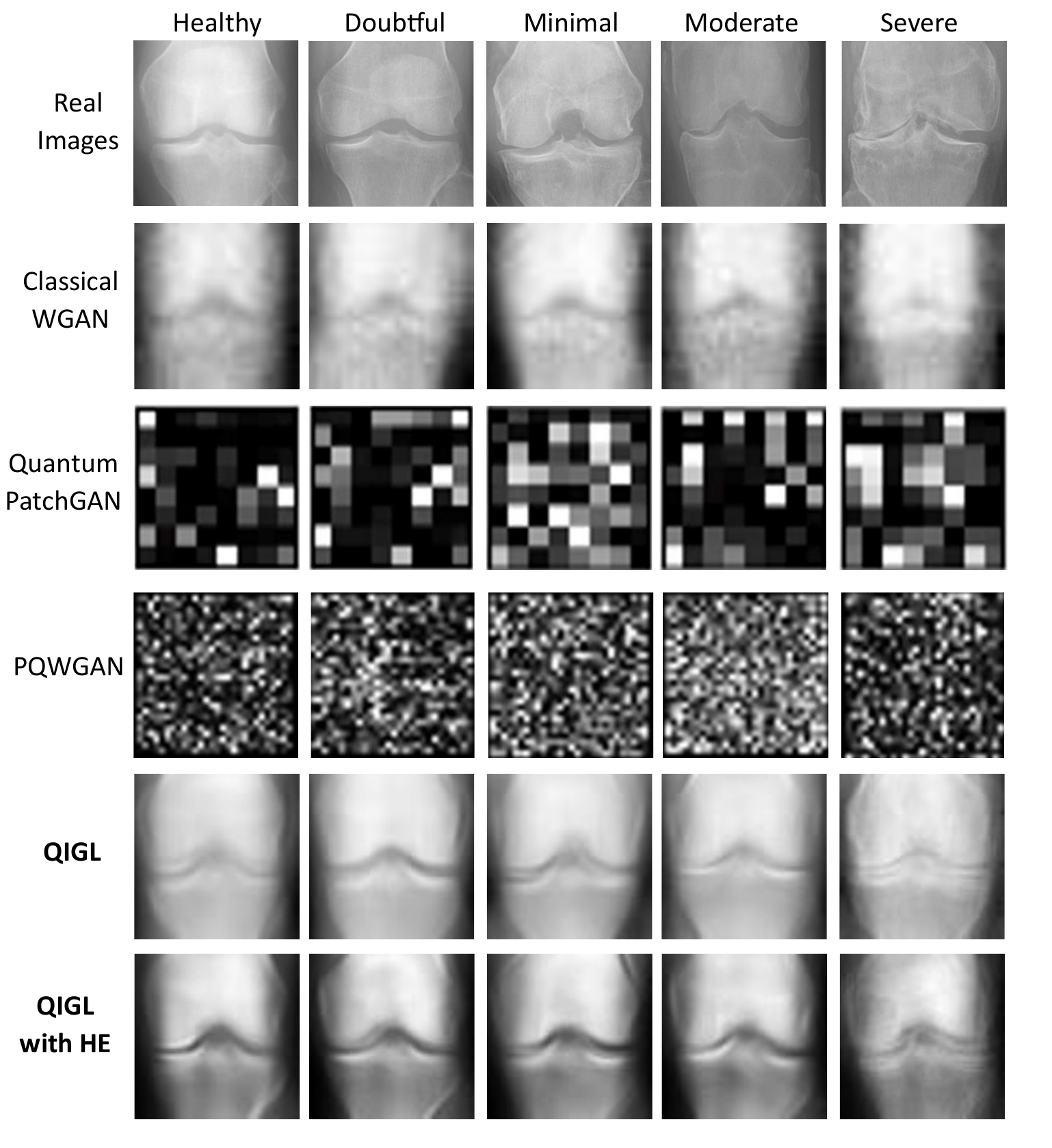}
\end{center}
\caption{Generated Images from Knee Osteoarthritis dataset. The first row shows the real images from all the classes. The second, third, fourth and fifth rows represent the images generated by classical WGAN \cite{arjovsky2017wasserstein}, Quantum PatchGAN \cite{huang2021experimental}, PQWGAN \cite{tsang2023hybrid}, the proposed QIGL, and QIGL with Histogram Equalisation.}
\label{fig:Generated_Knee}
\end{figure}

\begin{figure}
\begin{center}
\includegraphics[width=1.0\linewidth]{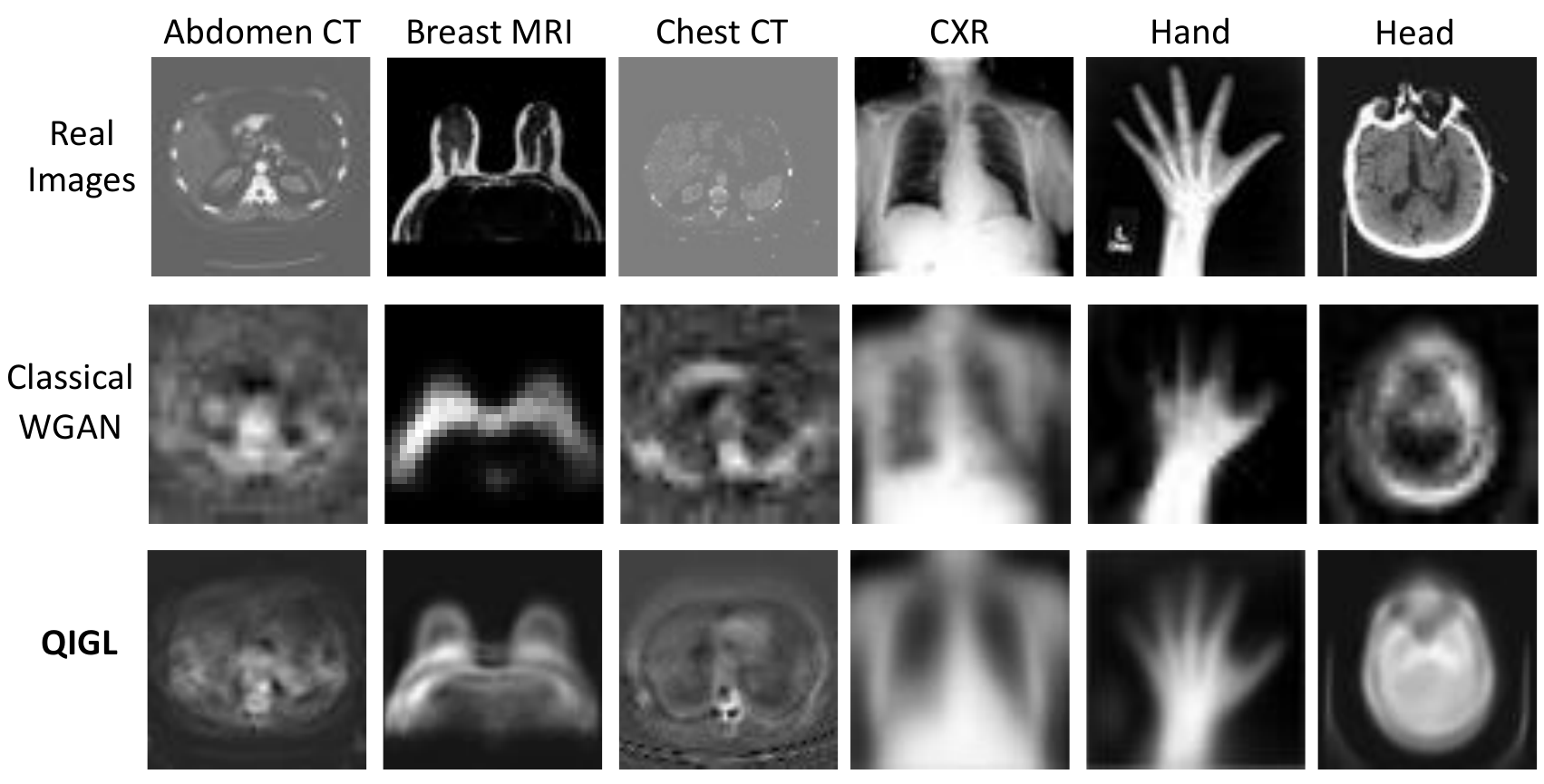}
\end{center}
\caption{Generated Images from medical MNIST dataset. The first row shows the real images from all the classes. For this dataset, we compare the generated images of the proposed QIGL with classical WGAN only.}
\label{fig:Generated_medical}
\end{figure}

\subsection{Qualitative Comparison with State-of-the-art Approaches}
A qualitative comparison of QGAN approaches is shown in Figure \ref{fig:Generated_Knee} for knee OA dataset. In the figure, the first row shows the real images for all classes. We implement classical WGAN \cite{arjovsky2017wasserstein}, Quantum PatchGAN \cite{huang2021experimental}, and PQWGAN \cite{tsang2023hybrid} for image generation and make a comparison with QIGL. For knee OA dataset, QIGL generates 10000 images, 2000 per class. The second, third, fourth and fifth row represent the generated images for classical WGAN, Quantum PatchGAN, PQWGAN, and our QIGL, respectively. The results indicate that, Quantum PatchGAN suffers from several challenges as it decomposes the image into patches and perform pixel-by-pixel learning, leading to difficulties in capturing global patterns and semantic details, particularly for high-resolution images. From the generated images, we find the quality of the images generated using Quantum PatchGAN is  low for knee OA data. While Quantum PatchGAN down-sized the MNIST hand-written digits data into $8 \times 8$ pixels, PQWGAN method \cite{tsang2023hybrid} generates images for the MNIST dataset with higher-resolution images ($28 \times 28$ pixels) compared to PatchGAN. Despite this promising performance of PQWGAN for MNIST dataset, it also incorporates the same patch strategy while integrating the concept of WGAN-GP (Wasserstein GAN-Gradient Penalty) \cite{gulrajani2017improved} to enhance convergence within the Quantum PatchGAN framework. Therefore, PQWGAN encounters challenges and fails to generate images for high-resolution datasets such as knee OA dataset  ($224 \times 224$ pixels). In contrast to these approaches and classical WGAN, QIGL generates high quality images as shown in Figure \ref{fig:Generated_Knee}. To further enhance the contrast of the images, we perform histogram equalisation (HE) as in Ref. \cite{prezja2022deepfake}, which is described in the METHOD \ref{Proposed Method} section. After HE, QIGL generates sharper images as illustrated in the last row.

We also compare QIGL generated images with the images generated using classical WGAN for medical MNIST dataset as depicted in Figure \ref{fig:Generated_medical}. We consider the same parameters for classical WGAN experiments as in QIGL for fair comparison which are described in Section \ref{Proposed Method}. From the figure, it is observed that our method generates images which are sharper and contain finer details than the images generated by the implemented classical WGAN. The performance of the proposed QIGL brings attention to the promising potential of quantum machine learning approaches for image generation tasks. While the comparison of QIGL and classical WGAN is insightful, it is important to note that WGAN is not considered state-of-the-art in GANs. Currently, StyleGAN \cite{karras2019style} and its variants \cite{karras2020analyzing, karras2021alias} are recognised for their superior performance in generating high-quality images with fine details and realistic textures. \textcolor{black}{The proposed approach is classically simulable. To achieve non-simulability, the the number of qubits and circuit depth in the quantum generator can be increased, although this raises concerns about barren plateaus. Mitigation techniques such as layer-wise optimisation \cite{skolik2021layerwise} and adaptive circuit pruning \cite{sim2021adaptive} can be explored. Furthermore, leveraging hardware-specific advantages, efficient data encoding methods, and hybrid quantum-classical architectures will allow scalability without compromising computational feasibility. However, quantum hardware are still in their infancy. The limited number of qubits available on current quantum computers, along with high error rates and decoherence, makes building complex models like StyleGAN infeasible at this stage. Nevertheless, our study has demonstrated that quantum generative models such as QIGL could enable high resolution image generation for medical applications and in future could be deployed on larger scale fault-tolerant quantum computers when they become available within the next decade.}

\subsection{Quantitative Comparison of QIGL with Classical WGAN}
\label{classicalWGAN}
To quantitatively compare the quality of images generated by QIGL with other methods, we calculate the Fréchet Inception Distance (FID) scores as an evaluation metric (see details in Section \ref{Proposed Method}) for both QIGL and classical WGAN on knee OA and medical MNIST datasets as reported in Table \ref{tab:QIGL_results}. The classical GANs \cite{heusel2017gans,zhao2020differentiable,karras2019style} utilise FID scores to evaluate the performance of generative models which is considered a standard method. FID score provides a quantitative measure of the similarity between the feature distributions of real and generated images, with lower values representing better performance. For FID score calculation, the feature representations are extracted for both real images and generated images using pre-trained deep convolutional neural network. Let us denote that the feature representations of real images as $\mu_r$ (mean) and $\Sigma_r$ (covariance) and the feature representations of generated images as $\mu_g$ (mean) and $\Sigma_g$ (covariance). The Fréchet distance between two multivariate Gaussian distributions $N(\mu_r, \Sigma_r)$ and $N(\mu_g, \Sigma_g)$ can be represented as,
\begin{align}
FID = \| \mu_r - \mu_g \|^2 + \text{Tr}(\Sigma_r + \Sigma_g - 2(\Sigma_r \Sigma_g)^{1/2}),
\end{align}
where $\text{Tr}(\cdot)$ denotes the trace operation which computes the sum of the diagonal elements of the resulting matrix and thus quantifies the dissimilarity between the covariance matrices. For each class of knee condition, the FID score obtained by both QIGL and classical WGAN method are listed in Table \ref{tab:QIGL_results}. For all the classes, FID score is lower for our method compared to the FID score obtained using images generated by classical WGAN, demonstrating that QIGL outperforms its classical counterpart across different levels of knee OA severity. We also calculated the FID score for the medical MNIST dataset. Similar to the case of Knee OA data, QIGL outperforms classical WGAN for all 6 classes. These results indicate superior performance of QIGL in generating realistic images across a diverse range of datasets when compared to the classical WGAN.

\begin{table*}[!t]
\centering
\small
\begin{tabular}{|p{2.5cm}|p{2.5cm}|p{2.5cm}|p{2.5cm}|p{2.5cm}|p{2.5cm}|}
\hline
\multicolumn{3}{|c|}{\textbf{Knee Osteoarthritis}} & \multicolumn{3}{c|}{\textbf{Medical MNIST}} \\
\hline \hline
\textbf{Classes} & \textbf{FID (QIGL)} & \textbf{FID (WGAN)} & \textbf{Classes} & \textbf{FID (QIGL)} & \textbf{FID (WGAN)} \\
\hline
\multicolumn{1}{|l|}{Healthy}   & 63.59 & 200.93 & AbdomenCT & 45.89 & 234.35 \\
\hline
\multicolumn{1}{|l|}{Doubtful}  & 56.28 & 188.29 & BreastMRI & 117.58 & 356.42 \\
\hline
\multicolumn{1}{|l|}{Minimal}   & 61.85 & 339.01 & ChestCT   & 42.95 & 212.83 \\
\hline
\multicolumn{1}{|l|}{Moderate}  & 62.76 & 312.58 & CXR       & 157.73 & 454.47 \\
\hline
\multicolumn{1}{|l|}{Severe}    & 75.54 & 310.07 & Hand      & 175.82 & 473.61 \\
\hline
 & & & \multicolumn{1}{|l|}{Head}  & 151.94 & 421.27       \\
\hline
\end{tabular}
\caption{Simulation results of the proposed QIGL approach compared to the classical WGAN method on Knee Osteoarthritis and Medical MNIST datasets. FID, representing the Fréchet Inception Distance, is the lower the better. }
\label{tab:QIGL_results}
\end{table*}

\begin{figure*}
    \centering
    \includegraphics[width=1.0\linewidth]{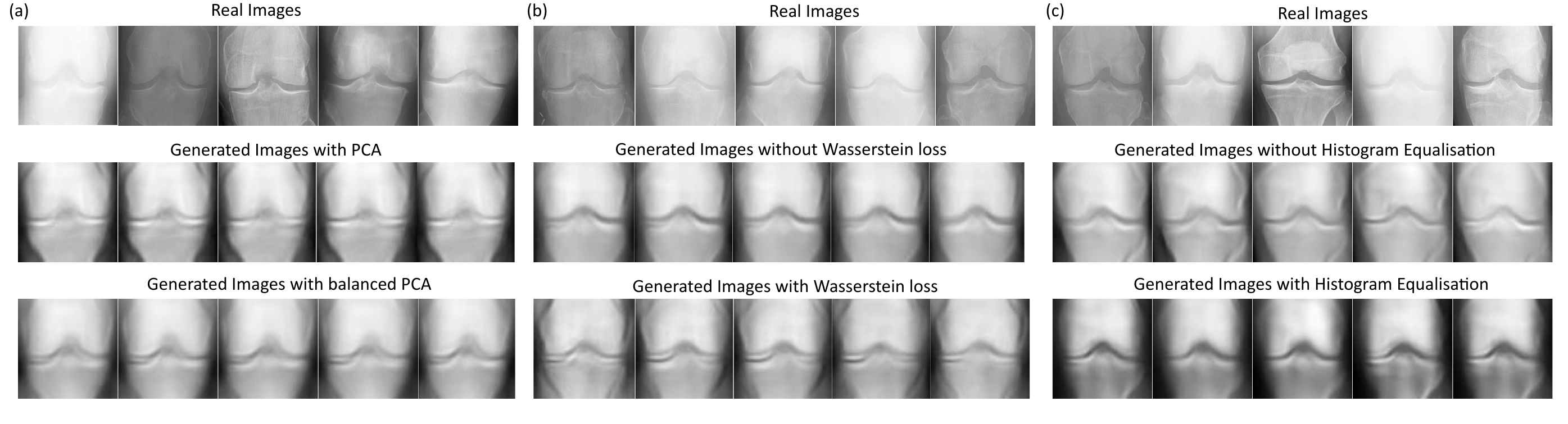}
    \caption{Generated images on Knee Osteoarthritis dataset
for class 0 (Healthy): (a) represents the generated images with conventional PCA feature distribution technique and balanced PCA feature distribution technique.
    (b) shows the generated images without Wasserstein loss and with Wasserstein loss, and (c) represents the generated images without Histogram Equalisation and with Histogram Equalisation.}
    \label{fig:Generated-PCA-WL-HE}
\end{figure*}

To study the training efficiency, we also calculated the number of parameters for both classical WGAN method (Table \ref{tab:WGANparameters} in the supplementary document) and QIGL (Table \ref{tab:QIGLparameters} in the supplementary document). Total number of training parameters in classical WGAN is 6,819,841 (4,718,592 for the generator and 2,101,249 for the discriminator) whereas the number of training parameters in QIGL is 3,489 (240 for the quantum generator and 3,249 for the discriminator). This suggests that the implemented quantum network offers better performance while using $\approx2000\times$ less number of training parameters. This exhibits the advantages of our QIGL approach in training efficiency by reducing the parameters significantly, highlighting its potency for practical implementation. The number of parameters in QIGL method can be increased by increasing the number of layers or by increasing the number of qubits in the quantum generator circuit. However, these improvements might cause well known barren plateau issue in which the optimisation landscape becomes flat. \textcolor{black}{This can impact training efficiency as gradient-based optimisers do not receive enough useful
direction for updates.}\textcolor{black}{For large images and models, alternative approaches can be explored, such as Quantum Convolutional Neural Networks (QCNNs), where a sequence of convolutional and pooling layers reduce the number of qubits while preserving information of relevant data features. Recent analysis \cite{pesah2021absence} demonstrates that QCNNs avoid barren plateaus, as the gradient variance diminishes at a polynomial rate rather than exponentially. Additionally, hybrid quantum-classical models can preprocess larger images by classical methods (e.g. downsampling or feature extraction) and reduce quantum circuit complexity. Layer wise learning \cite{skolik2021layerwise} can also
be explored to mitigate this issue as  sequential optimisation of layers ensures better gradient flow in deeper circuits. However, we leave increased the size problems as a future work.} Similarly, more advanced classical GAN models, such as StyleGAN \cite{karras2019style}, have been utilised in the literature for better quality medical image generation \cite{woodland2022evaluating, ahn2023high}. However, these advanced classical models require training of millions (~26 million for StyleGAN) of parameters and intensive computational resources. As we mitigate issues towards increasing problem size for our QIGL method, we will also study these advanced classical GAN methods for performance comparison of the classical and quantum generative methods with the goal of better quality generation of medical images.

\subsection{Ablation Studies}
We perform a set of simulations to evaluate the effectiveness of the proposed PCA feature re-distribution technique which is compared with traditional PCA feature distribution technique. We also conduct simulations to generate images with and without Wasserstein Distance, and with and without histogram equalisation to demonstrate the effect of these components. 
\begin{table}[!t]
\fontsize{8}{8}\selectfont
\centering
\begin{tabular}{|p{7cm}|p{1.1 cm}|}
\hline
\textbf{Method} & \textbf{FID}\\
\hline\hline
QIGL with PCA (no Wasserstein Loss) &81.4 \\
\hline
QIGL with PCA + Wasserstein Loss &78.1 \\
\hline
QIGL + Balanced PCA &76.8 \\
\hline
QIGL + Balanced PCA + Wasserstein Loss &71.6 \\
\hline
QIGL + Balanced PCA + Wasserstein Loss + HE &63.6 \\
\hline
\end{tabular}
\caption{Ablation studies on Knee Osteoarthritis dataset for class 0 (Healthy). ``QIGL with PCA'' is the proposed QIGL with normal PCA feature distribution, ``QIGL with PCA + Wasserstein Loss'' represents with normal PCA feature distribution but considering Wasserstein loss, ``QIGL + Balanced PCA'' is the method with balanced PCA feature distribution technique but there is no  Wasserstein loss, ``QIGL + Balanced PCA + Wasserstein Loss'' denotes the approach with both balanced PCA feature distribution technique and Wasserstein loss, and ``QIGL + Balanced PCA + Wasserstein Loss + HE'' represents the QIGL method with balanced PCA feature distribution technique, Wasserstein loss and histogram equalisation.}
\label{tab:ablation}
\end{table}

\subsubsection{Effect of PCA Balanced Feature Distribution}
In QIGL, PCA is considered to extract the principal features of images that facilitates efficient learning. Traditionally, PCA computes a set of principal components, which are linear combinations of the original features. These components are sequenced such that the first principal component captures the maximum amount of variance present in the data, the second principal component captures the second highest amount of variance, and so on. As there are eight sub-generators in the proposed framework as mentioned in Section \ref{generator_discriminator}, the PCA features are typically distributed among the sub-generators succeedingly. Hence, the initial sub-generators will get the first few features, which are heavily concentrated of the explained variance compared to the subsequent features. This sequential distribution leads to an imbalance in feature distribution among the sub-generators, potentially impacting the overall performance and diversity of the generated outputs. To overcome this problem, we incorporate a feature distribution technique to ensure that the features are distributed among the sub-generators in a more balanced way. This technique is explained in the supplementary document \ref{conventionalPCA} in more detail.

To investigate the effectiveness of the balanced distribution technique of PCA features, we train the QIGL model with normal feature distribution technique for comparison. Figure  \ref{fig:Generated-PCA-WL-HE}(a) shows the generated images on one of the classes (healthy i.e., class 0) from the knee OA dataset. While the top row shows the real images, the second and third rows depict the images generated with PCA conventional distribution and PCA balanced feature distribution, respectively. It is very hard to see the visual difference of the generated images for human perception. Therefore, we performed experiments to calculate the FID score of the generated images with both distributions of PCA features as in Table \ref{tab:ablation}. The FID score of QIGL with balanced PCA is 76.8 while the score is 81.4 with conventional PCA feature distribution.

\subsubsection{Effect of Wasserstein Distance in QIGL}
A limitation of our model is related to a tendency of generating repetitive and uniform images, thus lacking in variety and detail, which is a common issue even in classical GANs \cite{bang2021mggan, NIPS2017_44a2e080}. While it is essential to generate high-quality images, it is equally important to generate a diverse range of samples to enhance the model's fidelity to real-world data. We propose the integration of the Wasserstein distance (WD) into QIGL framework. WD encourages the generator to produce samples that captures the diversity present in the real data distribution. The impact of mode collapse and WD is thoroughly explained in the supplementary document \ref{mode-collapse}. Figure  \ref{fig:Generated-PCA-WL-HE}(b) illustrates the generated images both with and without the Wasserstein loss for class 1 of Knee OA dataset. From the Figure, it can be observed that the model exhibits enhanced diversity in samples when WD is used. We further evaluate the effectiveness of WD by calculating the FID score as shown in Table \ref{tab:ablation}. From the Table, the QIGL achieves an FID score of 81.4 when there is no WD, which is 78.1 when WD is incorporating within QIGL, demonstrating the effectiveness of this approach in mitigating mode collapse and enhancing image generation diversity.

\subsubsection{Effect of Histogram Equalisation}
In this subsection, we investigate the impact of histogram equalisation (HE) on the quality of generated knee X-ray images. Given that real medical X-ray images used for generating synthetic images often lack sharpness, we apply HE to enhance the contrast of these real input images. Subsequently, we assess the effect of this pre-processing step on the fidelity of the generated samples and corresponding FID scores. Figure \ref{fig:Generated-PCA-WL-HE}(c) shows higher contrast in the generated images with HE compared to the images when HE is not considered. As a result of this, the FID score of the QIGL generated images including HE is lower for all classes of the knee images. We present the mean FID score for each class with and without HE implemented in Table \ref{tab:ablationHE} in the supplementary document. We observe that the HE procedure significantly improves FID score values.

\section{Discussion}
\textcolor{black}{The proposed QIGL addresses the challenges of the current QGAN method \cite{huang2021experimental}. While Ref. \cite{huang2021experimental} divides full images into smaller patches due to the limited number of qubits available in the current quantum devices and perform pixel-wise learning,  often leading to a loss of global structure and semantic information, results in poor quality image generation. In contrast, our approach uses PCA to extract the most important global features of the images, capturing the underlying structure without the dividing the images into patches. By selecting the top 40 principal components, we retain 98\% of the variability in the images, thus preserving global information. These components are then distributed across multiple sub-generators in the quantum circuit. This method addresses the scalability issues associated with patch-based approaches, while maintaining the integrity of the global structure of images. while the generator utilises multiple sub-generators, it does not rely on patch-based learning of local regions of the image. Instead, it focuses on global feature extraction through PCA, which significantly improves the quality of the generated images.}

The proposed method successfully generates high-quality medical images ($64 \times 64$ pixel) as shown in Figure \ref{fig:Generated_Knee} and \ref{fig:Generated_medical}.Through qualitative comparison with classical WGAN and existing QGAN approaches such as Quantum PatchGAN and PQWGAN, it is observed that existing methods struggle with capturing global patterns and semantic details, particularly in high-resolution datasets like the knee OA.  Our QIGL framework consistently generates higher quality images with sharper details. Integration of balanced PCA feature distribution technique effectively distributes PCA features across the sub-generators compare to the traditional PCA methods as shown in Figure \ref{fig:PCA-mode-collapse}(b) and Table \ref{tab:ablation}. Additionally, the incorporation of WD as a loss function within the proposed framework addresses mode collapse, leading to more diverse and realistic medical image generation. The quantitative comparisons through FID scores confirm the superiority of our approach over classical WGAN across various classes and datasets, highlighting the potential of quantum machine learning approaches in medical imaging tasks for improving healthcare outcomes through accurate diagnosis and analysis of medical images. However, quantum hardware is not yet scaled up to process complex quantum circuits which are necessary for the wide spread implementation of QGANs for real-world applications, and it will take a few more years to deploy such techniques on real devices.

\textcolor{black}{However, when transitioning to real quantum devices, several challenges may arise, such as efficient gradient estimation, and barren plateaus. As quantum circuits use parameterised quantum gates, gradients need to be computed during optimisation. However, on real quantum devices, gradient estimation is computationally expensive due to finite sampling noise and hardware constraints. To improve efficiency of gradient estimation, the gradients can be computed by parameter-shift rule \cite{wierichs2022general}, which requires evaluating the circuit at two shifted values of the parameter. While this method is exact, two circuit evaluations per parameter, which can be costly on real devices.
To improve efficiency, the parameter-shift rule can be replaced with Simultaneous Perturbation Stochastic Approximation (SPSA) \cite{periyasamy2024guided}, which estimates gradients using only two measurements per optimization step, irrespective of the number of parameters, making it more practical for large-scale training. Further efficiency gains can be achieved through adaptive measurement protocols \cite{kubler2020adaptive}, which dynamically allocate measurement resources based on training progress, focusing computational effort where it is most needed. These techniques might help optimise measurement efficiency and improve gradient estimation accuracy on real quantum hardware.}

 \textcolor{black}{Moreover, when deploying on real quantum  hardware, QIGL’s generator can interact directly with a classical discriminator in an adversarial learning framework, where the quantum generator produces measured expectation values that are evaluated classically. This eliminates the need for classical simulation of quantum states while maintaining a feasible training loop. To further reduce quantum circuit complexity, Quantum Convolutional Neural Networks (QCNNs) and layer-wise learning can be employed, ensuring that gradient magnitudes remain significant even as the model scales. While these strategies improve the feasibility of training QIGL on real quantum hardware, challenges such as optimizing deep quantum models and improving the efficiency of hardware-based training remain open problems. We leave these as future research directions, and our work on QIGL provides a foundation for exploring scalable quantum generative learning models, enabling further advancements in quantum-enhanced image synthesis and optimization techniques.}

\section{Conclusion and Future Work}
In this paper, we presented QIGL framework for high-quality medical image generation that significantly advances the application of quantum generative learning in healthcare. Through a comprehensive set of simulations and analysis  on a diverse set of classes of medical image datasets, we have demonstrated superior performance of our approach compared to its classical counterparts and the existing QGAN methods, particularly in terms of image quality, scalability, mode collapse mitigation, and resource efficiency. Our simulation results and extensive comparison of QIGL with its classical counterparts open up avenues for future
research in healthcare technology. While our results demonstrate remarkable gains in medical image generation, there remain opportunities for further exploration, such as extending QIGL approach for RGB medical image generation, implementation on real quantum devices, and perform image classification tasks, thus expanding the scope and applicability of quantum-enhanced techniques in medical imaging.

\section*{Acknowledgments}
A.K. acknowledge the use of CSIRO HPC (High-Performance Computing) for conducting the experiments. A.K. also acknowledges CSIRO's Quantum Technologies Future Science Platform for providing the opportunity to work on quantum machine learning. KYA and YSW were supported through Quantum Horizon Program at MITRE and MITRE National Security (MNS) Sector. \copyright 2024 The MITRE Corporation.
ALL RIGHTS RESERVED. Approved for public release. Distribution unlimited PR\textunderscore24$-$00320$-$4.

\bibliographystyle{unsrt} 
\bibliography{apssamp}% Produces the bibliography via BibTeX.

\begin{thebibliography}{10}

\bibitem{esteva2017dermatologist}
Andre Esteva, Brett Kuprel, Roberto~A Novoa, Justin Ko, Susan~M Swetter, Helen~M Blau, and Sebastian Thrun.
\newblock Dermatologist-level classification of skin cancer with deep neural networks.
\newblock {\em nature}, 542(7639):115--118, 2017.

\bibitem{litjens2017survey}
Geert Litjens, Thijs Kooi, Babak~Ehteshami Bejnordi, Arnaud Arindra~Adiyoso Setio, Francesco Ciompi, Mohsen Ghafoorian, Jeroen~Awm Van Der~Laak, Bram Van~Ginneken, and Clara~I S{\'a}nchez.
\newblock A survey on deep learning in medical image analysis.
\newblock {\em Medical image analysis}, 42:60--88, 2017.

\bibitem{stokes2020deep}
Jonathan~M Stokes, Kevin Yang, Kyle Swanson, Wengong Jin, Andres Cubillos-Ruiz, Nina~M Donghia, Craig~R MacNair, Shawn French, Lindsey~A Carfrae, Zohar Bloom-Ackermann, et~al.
\newblock A deep learning approach to antibiotic discovery.
\newblock {\em Cell}, 180(4):688--702, 2020.

\bibitem{goodfellow2014generative}
Ian Goodfellow, Jean Pouget-Abadie, Mehdi Mirza, Bing Xu, David Warde-Farley, Sherjil Ozair, Aaron Courville, and Yoshua Bengio.
\newblock Generative adversarial nets.
\newblock {\em Advances in neural information processing systems}, 27, 2014.

\bibitem{khatun2021end}
Amena Khatun, Simon Denman, Sridha Sridharan, and Clinton Fookes.
\newblock End-to-end domain adaptive attention network for cross-domain person re-identification.
\newblock {\em IEEE Transactions on Information Forensics and Security}, 16:3803--3813, 2021.

\bibitem{khatun2023pose}
Amena Khatun, Simon Denman, Sridha Sridharan, and Clinton Fookes.
\newblock Pose-driven attention-guided image generation for person re-identification.
\newblock {\em Pattern Recognition}, 137:109246, 2023.

\bibitem{karras2020analyzing}
Tero Karras, Samuli Laine, Miika Aittala, Janne Hellsten, Jaakko Lehtinen, and Timo Aila.
\newblock Analyzing and improving the image quality of {S}tyle{GAN}.
\newblock In {\em Proceedings of the IEEE/CVF conference on computer vision and pattern recognition}, pages 8110--8119, 2020.

\bibitem{brock2018large}
Andrew Brock, Jeff Donahue, and Karen Simonyan.
\newblock Large scale {GAN} training for high fidelity natural image synthesis.
\newblock {\em arXiv preprint arXiv:1809.11096}, 2018.

\bibitem{iqbal2022survey}
Touseef Iqbal and Shaima Qureshi.
\newblock The survey: Text generation models in deep learning.
\newblock {\em Journal of King Saud University-Computer and Information Sciences}, 34(6):2515--2528, 2022.

\bibitem{Radford2015UnsupervisedRL}
Alec Radford, Luke Metz, and Soumith Chintala.
\newblock Unsupervised representation learning with deep convolutional generative adversarial networks.
\newblock {\em CoRR}, abs/1511.06434, 2015.

\bibitem{pang2021deep}
Guansong Pang, Chunhua Shen, Longbing Cao, and Anton Van~Den Hengel.
\newblock Deep learning for anomaly detection: A review.
\newblock {\em ACM computing surveys (CSUR)}, 54(2):1--38, 2021.

\bibitem{karras2019style}
Tero Karras, Samuli Laine, and Timo Aila.
\newblock A style-based generator architecture for generative adversarial networks.
\newblock In {\em Proceedings of the IEEE/CVF conference on computer vision and pattern recognition}, pages 4401--4410, 2019.

\bibitem{lloyd2018quantum}
Seth Lloyd and Christian Weedbrook.
\newblock Quantum generative adversarial learning.
\newblock {\em Physical review letters}, 121(4):040502, 2018.

\bibitem{dallaire2018quantum}
Pierre-Luc Dallaire-Demers and Nathan Killoran.
\newblock Quantum generative adversarial networks.
\newblock {\em Physical Review A}, 98(1):012324, 2018.

\bibitem{nakaji2021quantum}
Kouhei Nakaji and Naoki Yamamoto.
\newblock Quantum semi-supervised generative adversarial network for enhanced data classification.
\newblock {\em Scientific reports}, 11(1):19649, 2021.

\bibitem{huang2021experimental}
He-Liang Huang, Yuxuan Du, Ming Gong, Youwei Zhao, Yulin Wu, Chaoyue Wang, Shaowei Li, Futian Liang, Jin Lin, Yu~Xu, et~al.
\newblock Experimental quantum generative adversarial networks for image generation.
\newblock {\em Physical Review Applied}, 16(2):024051, 2021.

\bibitem{tsang2023hybrid}
Shu~Lok Tsang, Maxwell~T West, Sarah~M Erfani, and Muhammad Usman.
\newblock Hybrid quantum-classical generative adversarial network for high resolution image generation.
\newblock {\em IEEE Transactions on Quantum Engineering}, 2023.

\bibitem{chen2018knee}
Pingjun Chen.
\newblock Knee osteoarthritis severity grading dataset.
\newblock {\em Mendeley Data}, 1:21--23, 2018.

\bibitem{9098456}
Bofei Zhang, Jimin Tan, Kyunghyun Cho, Gregory Chang, and Cem~M. Deniz.
\newblock Attention-based {CNN} for {KL} grade classification: Data from the osteoarthritis initiative.
\newblock In {\em 2020 IEEE 17th International Symposium on Biomedical Imaging (ISBI)}, pages 731--735, 2020.

\bibitem{arjovsky2017wasserstein}
Martin Arjovsky, Soumith Chintala, and L{\'e}on Bottou.
\newblock Wasserstein generative adversarial networks.
\newblock In {\em International conference on machine learning}, pages 214--223. PMLR, 2017.

\bibitem{khatun2020semantic}
Amena Khatun, Simon Denman, Sridha Sridharan, and Clinton Fookes.
\newblock Semantic consistency and identity mapping multi-component generative adversarial network for person re-identification.
\newblock In {\em Proceedings of the IEEE/CVF Winter Conference on Applications of Computer Vision}, pages 2267--2276, 2020.

\bibitem{gatys2016image}
Leon~A. Gatys, Alexander~S. Ecker, and Matthias Bethge.
\newblock Image style transfer using convolutional neural networks.
\newblock In {\em Proceedings of the IEEE conference on computer vision and pattern recognition}, pages 2414--2423, 2016.

\bibitem{yoo2016pixel}
Donggeun Yoo, Namil Kim, Sunggyun Park, Anthony~S. Paek, and In~So Kweon.
\newblock Pixel-level domain transfer.
\newblock In {\em Computer Vision--ECCV 2016: 14th European Conference, Amsterdam, The Netherlands, October 11-14, 2016, Proceedings, Part VIII 14}, pages 517--532. Springer, 2016.

\bibitem{heusel2017gans}
Martin Heusel, Hubert Ramsauer, Thomas Unterthiner, Bernhard Nessler, and Sepp Hochreiter.
\newblock Gans trained by a two time-scale update rule converge to a local nash equilibrium.
\newblock {\em Advances in neural information processing systems}, 30, 2017.

\bibitem{qu2023quantum}
Zhiguo Qu, Wenke Shi, and Prayag Tiwari.
\newblock Quantum conditional generative adversarial network based on patch method for abnormal electrocardiogram generation.
\newblock {\em Computers in Biology and Medicine}, 166:107549, 2023.

\bibitem{zhou2023hybrid}
Nan-Run Zhou, Tian-Feng Zhang, Xin-Wen Xie, and Jun-Yun Wu.
\newblock Hybrid quantum--classical generative adversarial networks for image generation via learning discrete distribution.
\newblock {\em Signal Processing: Image Communication}, 110:116891, 2023.

\bibitem{silver2023mosaiq}
Daniel Silver, Tirthak Patel, William Cutler, Aditya Ranjan, Harshitta Gandhi, and Devesh Tiwari.
\newblock Mosaiq: Quantum generative adversarial networks for image generation on nisq computers.
\newblock In {\em Proceedings of the IEEE/CVF International Conference on Computer Vision}, pages 7030--7039, 2023.

\bibitem{landman2022quantum}
Jonas Landman, Natansh Mathur, Yun~Yvonna Li, Martin Strahm, Skander Kazdaghli, Anupam Prakash, and Iordanis Kerenidis.
\newblock Quantum methods for neural networks and application to medical image classification.
\newblock {\em Quantum}, 6:881, 2022.

\bibitem{cherrat2024quantum}
El~Amine Cherrat, Iordanis Kerenidis, Natansh Mathur, Jonas Landman, Martin Strahm, and Yun~Yvonna Li.
\newblock Quantum vision transformers.
\newblock {\em Quantum}, 8(arXiv: 2209.08167):1265, 2024.

\bibitem{stein2021qugan}
Samuel~A Stein, Betis Baheri, Daniel Chen, Ying Mao, Qiang Guan, Ang Li, Bo~Fang, and Shuai Xu.
\newblock Qugan: A quantum state fidelity based generative adversarial network.
\newblock In {\em 2021 IEEE International Conference on Quantum Computing and Engineering (QCE)}, pages 71--81. IEEE, 2021.

\bibitem{paszke2019pytorch}
Adam Paszke, Sam Gross, Francisco Massa, Adam Lerer, James Bradbury, Gregory Chanan, Trevor Killeen, Zeming Lin, Natalia Gimelshein, Luca Antiga, et~al.
\newblock Pytorch: An imperative style, high-performance deep learning library.
\newblock {\em Advances in neural information processing systems}, 32, 2019.

\bibitem{bergholm2018pennylane}
Ville Bergholm, Josh Izaac, Maria Schuld, Christian Gogolin, Shahnawaz Ahmed, Vishnu Ajith, M~Sohaib Alam, Guillermo Alonso-Linaje, B~AkashNarayanan, Ali Asadi, et~al.
\newblock Pennylane: Automatic differentiation of hybrid quantum-classical computations.
\newblock {\em arXiv preprint arXiv:1811.04968}, 2018.

\bibitem{kingma2014adam}
Diederik~P. Kingma and Jimmy Ba.
\newblock Adam: A method for stochastic optimization.
\newblock {\em arXiv preprint arXiv:1412.6980}, 2014.

\bibitem{prezja2022deepfake}
Fabi Prezja, Juha Paloneva, Ilkka P{\"o}l{\"o}nen, Esko Niinim{\"a}ki, and Sami {\"A}yr{\"a}m{\"o}.
\newblock Deep{F}ake knee osteoarthritis {X}-rays from generative adversarial neural networks deceive medical experts and offer augmentation potential to automatic classification.
\newblock {\em Scientific Reports}, 12(1):18573, 2022.

\bibitem{MedicalMNISTClassification}
Medical mnist dataset (available at: https://www.kaggle.com/datasets/andrewmvd/medical-mnist).

\bibitem{gulrajani2017improved}
Ishaan Gulrajani, Faruk Ahmed, Martin Arjovsky, Vincent Dumoulin, and Aaron~C Courville.
\newblock Improved training of {W}asserstein {GAN}s.
\newblock {\em Advances in neural information processing systems}, 30, 2017.

\bibitem{karras2021alias}
Tero Karras, Miika Aittala, Samuli Laine, Erik H{\"a}rk{\"o}nen, Janne Hellsten, Jaakko Lehtinen, and Timo Aila.
\newblock Alias-free generative adversarial networks.
\newblock {\em Advances in neural information processing systems}, 34:852--863, 2021.

\bibitem{skolik2021layerwise}
Andrea Skolik, Jarrod~R McClean, Masoud Mohseni, Patrick Van Der~Smagt, and Martin Leib.
\newblock Layerwise learning for quantum neural networks.
\newblock {\em Quantum Machine Intelligence}, 3:1--11, 2021.

\bibitem{sim2021adaptive}
Sukin Sim, Jonathan Romero, J{\'e}r{\^o}me~F Gonthier, and Alexander~A Kunitsa.
\newblock Adaptive pruning-based optimization of parameterized quantum circuits.
\newblock {\em Quantum Science and Technology}, 6(2):025019, 2021.

\bibitem{zhao2020differentiable}
Shengyu Zhao, Zhijian Liu, Ji~Lin, Jun-Yan Zhu, and Song Han.
\newblock Differentiable augmentation for data-efficient gan training.
\newblock {\em Advances in neural information processing systems}, 33:7559--7570, 2020.

\bibitem{pesah2021absence}
Arthur Pesah, Marco Cerezo, Samson Wang, Tyler Volkoff, Andrew~T Sornborger, and Patrick~J Coles.
\newblock Absence of barren plateaus in quantum convolutional neural networks.
\newblock {\em Physical Review X}, 11(4):041011, 2021.

\bibitem{woodland2022evaluating}
McKell Woodland, John Wood, Brian~M Anderson, Suprateek Kundu, Ethan Lin, Eugene Koay, Bruno Odisio, Caroline Chung, Hyunseon~Christine Kang, Aradhana~M Venkatesan, et~al.
\newblock Evaluating the performance of stylegan2-ada on medical images.
\newblock In {\em International Workshop on Simulation and Synthesis in Medical Imaging}, pages 142--153. Springer, 2022.

\bibitem{ahn2023high}
Gun Ahn, Byung~Sun Choi, Sunho Ko, Changwung Jo, Hyuk-Soo Han, Myung~Chul Lee, and Du~Hyun Ro.
\newblock High-resolution knee plain radiography image synthesis using style generative adversarial network adaptive discriminator augmentation.
\newblock {\em Journal of Orthopaedic Research{\textregistered}}, 41(1):84--93, 2023.

\bibitem{bang2021mggan}
Duhyeon Bang and Hyunjung Shim.
\newblock Mggan: Solving mode collapse using manifold-guided training.
\newblock In {\em Proceedings of the IEEE/CVF international conference on computer vision}, pages 2347--2356, 2021.

\bibitem{NIPS2017_44a2e080}
Akash Srivastava, Lazar Valkov, Chris Russell, Michael~U. Gutmann, and Charles Sutton.
\newblock Veegan: Reducing mode collapse in gans using implicit variational learning.
\newblock In I.~Guyon, U.~Von Luxburg, S.~Bengio, H.~Wallach, R.~Fergus, S.~Vishwanathan, and R.~Garnett, editors, {\em Advances in Neural Information Processing Systems}, volume~30. Curran Associates, Inc., 2017.

\bibitem{wierichs2022general}
David Wierichs, Josh Izaac, Cody Wang, and Cedric Yen-Yu Lin.
\newblock General parameter-shift rules for quantum gradients.
\newblock {\em Quantum}, 6:677, 2022.

\bibitem{periyasamy2024guided}
Maniraman Periyasamy, Axel Plinge, Christopher Mutschler, Daniel~D Scherer, and Wolfgang Mauerer.
\newblock Guided-spsa: Simultaneous perturbation stochastic approximation assisted by the parameter shift rule.
\newblock In {\em 2024 IEEE International Conference on Quantum Computing and Engineering (QCE)}, volume~1, pages 1504--1515. IEEE, 2024.

\bibitem{kubler2020adaptive}
Jonas~M K{\"u}bler, Andrew Arrasmith, Lukasz Cincio, and Patrick~J Coles.
\newblock An adaptive optimizer for measurement-frugal variational algorithms.
\newblock {\em Quantum}, 4:263, 2020.

\end{thebibliography}
%\end{comment}
%---------------------------------------------------------------
\clearpage

% Customizing figure numbering format and caption name

%\renewcommand{\thefigure}{S\arabic{figure}} 
%\renewcommand{\thetable}{S\arabic{table}}  
\makeatletter
\renewcommand \thesection{S\@arabic\c@section}
\renewcommand\thetable{S\@arabic\c@table}
\renewcommand \thefigure{S\@arabic\c@figure}
\makeatother

\begin{center}
   {\Large{\textbf{Supplementary Document}}}
\end{center}

\renewcommand{\thesection}{S\arabic{section}} % Prefix with 'S'
\setcounter{section}{0}

\renewcommand{\thefigure}{S\arabic{figure}}  % Change format globally for supplements
\setcounter{figure}{0}

\begin{figure*}
    \includegraphics[width=0.9\linewidth]{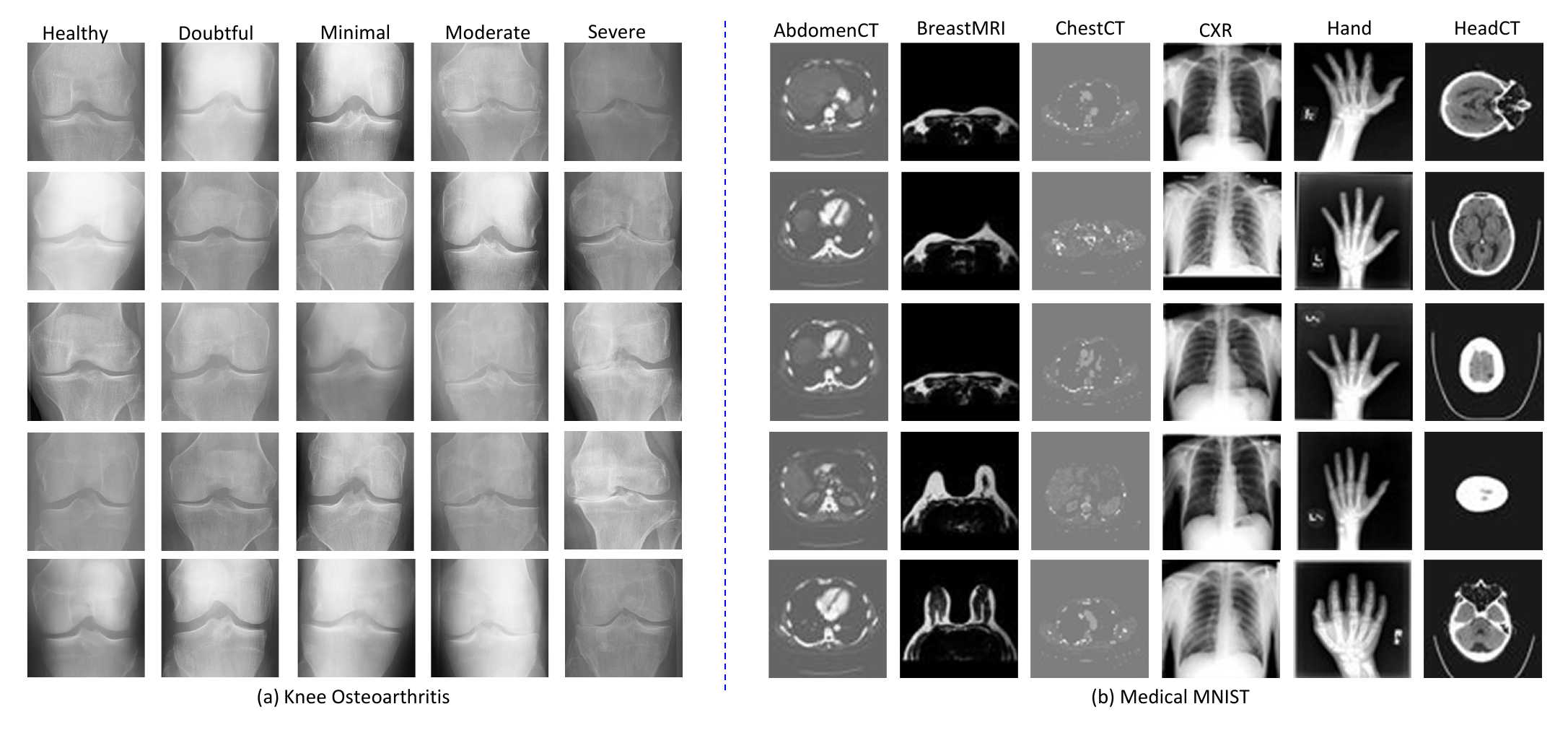}
    \caption{Overview of datasets used to generate synthetic images by the proposed QIGL. In the Figure, (a) represents Knee Osteoarthritis dataset, containing 5 classes, and (b) represents Medical MNIST dataset consisting of 6 classes.}
    \label{fig:Datasets}
\end{figure*}

\begin{figure*}
    \includegraphics[width=0.7\linewidth]{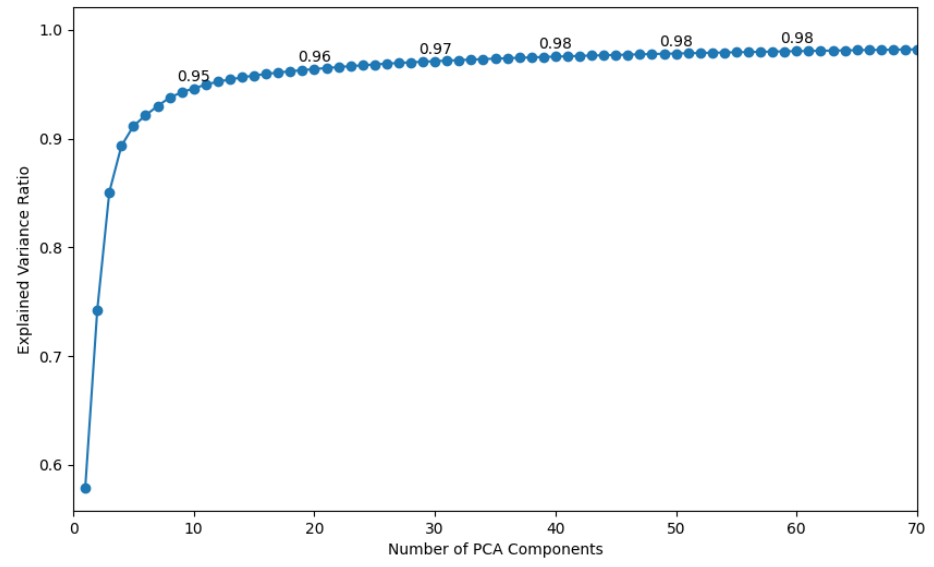}
    \caption{Plot of the explained variance ratio against the number of components. Here, 98\% of the information are captured by 40 principal components.}
    \label{fig:scree-plot}
\end{figure*}

\begin{figure*}
\begin{center}
\includegraphics[width=0.9\linewidth]{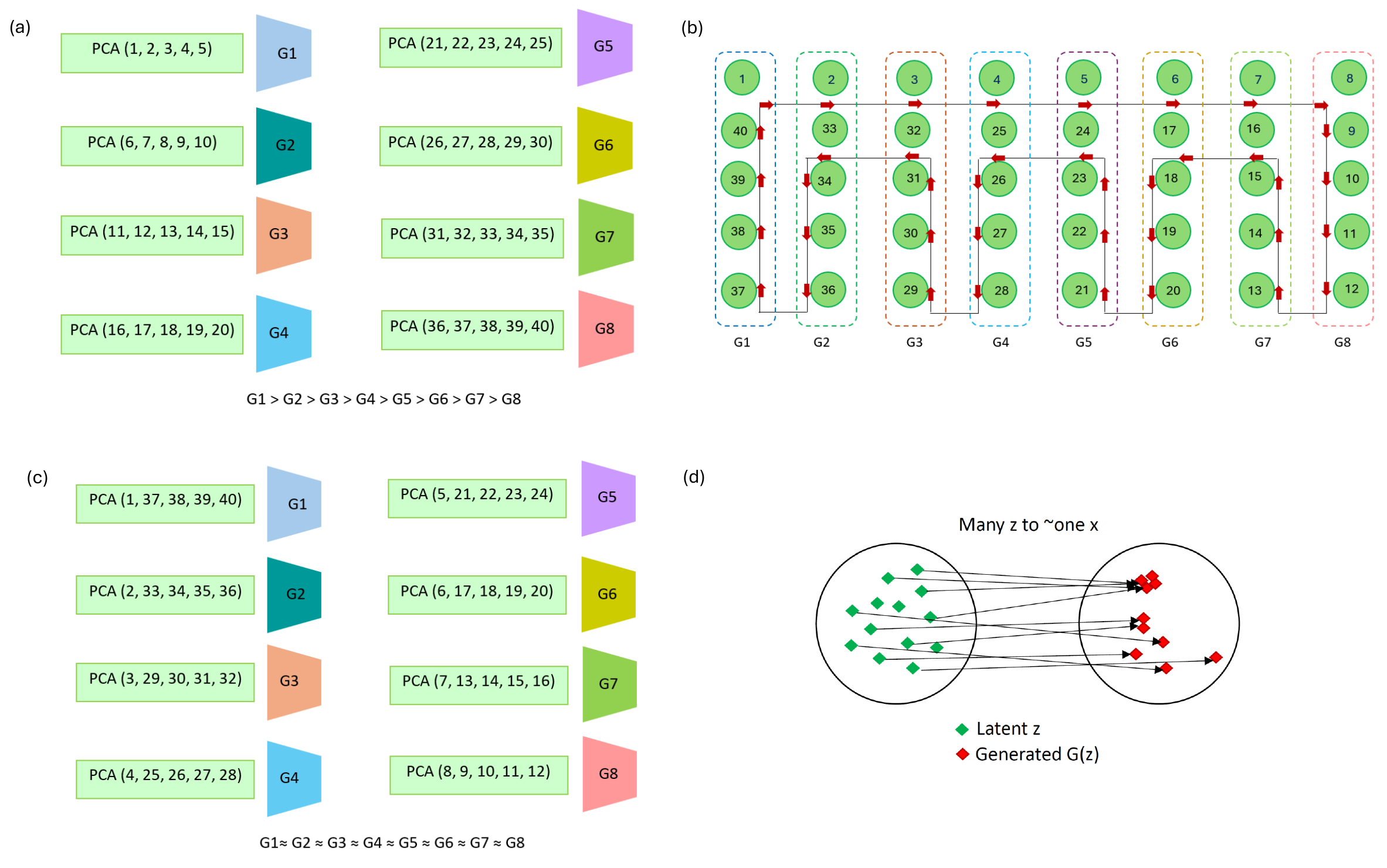}
\end{center}
\caption{Illustration of (a) conventional PCA feature distribution technique. QIGL consists of 8 sub-generators that requires the extraction of 40 PCA features for a 5-qubit quantum circuit. Each sub-generator is allocated with 5 PCA features.
The allocation of PCA features follows a systematic pattern: the first 5 features are directed to the first sub-generator (G1), the subsequent 5 features are allocated to the second sub-generator (G2), and the final 5 PCA features are directed to the eighth sub-generator (G8). (b) Balanced PCA feature re-distribution technique where the top principal component are assigned to the first sub-generator. Succeeding components are then sequentially allocated to each sub-generator until each holds one top principal component. Following this, we adopt a reverse distribution approach for the remaining principal components. This entails assigning the last $n-1$ components to the first sub-generator, the second-to-last $n-1$ components to the second sub-generator, and so forth, until all features are allocated principal component. (c) Balanced PCA feature distribution across the sub-generators, and (d) Mode Collapse problem where multiple latent distributions $G(z)$ is mapped to only one real space output $x$, resulting limited set of output samples from generator.}
\label{fig:PCA-mode-collapse}
\end{figure*}

\begin{table*}
\centering
\begin{tabular}{|p{3.2cm}|p{3cm}|p{3cm}|}
\hline
\textbf{Layer} & \textbf{Generator} & \textbf{Discriminator}\\
\hline\hline
Layer 1 (weights) & 100 $\times$ 128 & (1$\times$ 64 $\times$ 64) $\times$ 512 \\
\hline
Layer 1 (biases) & 128 & 512 \\
\hline
Layer 2 (weights)  &128 $\times$ 256 & 512 $\times$ 256\\
\hline
Layer 2 (biases)  & 256 & 256 \\
\hline
Layer 3 (weights)  &256 $\times$ 512 & 256 $\times$ 1\\
\hline
Layer 3 (biases)  & 512 & 1 \\
\hline
Layer 4 (weights)  & 512 $\times$ 1024 & -  \\
\hline
Layer 4 (biases)  &1024 &- \\
\hline
Layer 5 (weights) &1024 $\times$ (1$\times$ 64 $\times$ 64) &- \\
\hline
Layer 5 (biases) & (1$\times$ 64 $\times$ 64) &- \\
\hline
Batch normalization &256+512+1024 &- \\
\hline
{\bf{Total}} & 4,718,592 & 2,101,249\\
\hline
\end{tabular}
\caption{Number of training parameters at each layer of generator and discriminator for classical WGAN.}
\label{tab:WGANparameters}
\end{table*}

\begin{table*}
\centering
\begin{tabular}{|p{4cm}|p{2.5cm}|}
\hline
\textbf{Layer} & \textbf{Discriminator} \\
\hline\hline
Layer 1 (weights) & 40 $\times$ 64  \\
\hline
Layer 1 (biases) & 64  \\
\hline
Layer 2 (weights)  &64 $\times$ 16 \\
\hline
Layer 2 (biases)  & 16  \\
\hline
Layer 3 (weights)  &16 $\times$ 1 \\
\hline
Layer 3 (biases)  & 1 \\
\hline
\textbf{Total} & 3249 \\
\hline
\hline
\textbf{Quantum Generator} & 8 $\times$ 6 $\times$ 5 =240 \\
\hline
\end{tabular}
\caption{Number of training parameters at each layer of quantum generator and classical discriminator for QIGL.}
\label{tab:QIGLparameters}
\end{table*}

\begin{table*}
\centering
\begin{tabular}{|p{1.0cm}|p{1.2 cm}|p{2.8 cm}|}
\hline
\textbf{Class} & \textbf{QIGL} & \textbf{QIGL $+$ HE}\\
\hline\hline
0 & 67.50 & 63.59\\
\hline
1 & 63.18 & 56.28\\
\hline
2 & 73.76 & 61.85 \\
\hline
3 & 67.89 & 62.76\\
\hline
4 & 88.76 & 75.54 \\
\hline
\end{tabular}
\caption{The mean FID score values of the QIGL generated images without histogram equalization and with histogram equalization after 20 epochs for knee X-ray images dataset.}
\label{tab:ablationHE}
\end{table*}
\begin{table*}[htbp]
\centering
    \fontsize{8}{8}\selectfont
    \centering
    \begin{tabular}{|m{1.8cm}|m{3.1cm}|m{3.1cm}|}
    \hline
    \centering{\bf{Class}} & \centering{\bf{Number of damaged items}} & \centering\arraybackslash{\bf{Number of negative channel images}}\\
    \hline
    \hline
    \centering 0 & \centering 11 & \centering\arraybackslash 26\\
    \hline
    \centering 1 & \centering 5 & \centering\arraybackslash 8\\
    \hline
    \centering 2 & \centering 11 & \centering\arraybackslash 22\\
    \hline
    \centering 3 & \centering 12 & \centering\arraybackslash 10\\
    \hline
    \centering 4 & \centering 5 & \centering\arraybackslash 0\\
    \hline
\end{tabular}
\caption{The number of images with surgical prosthetics or scratches and punch holes (middle column) and number of negative channel images (last column) in the knee X-ray images dataset listed for each class, i.e., KL grade as described in the main manuscript.}
\label{tab:preprocessing}
\end{table*}

\section{Classical Generative Adversarial Networks (CGANs)}
\label{Sup-classicalGAN}
GANs are neural networks that learn to draw samples from a distribution they have learned by setting up a competition. The discriminator tries to distinguish whether the samples are real or generated, and the generator generates fake samples and tries to fool the discriminator. Thus, the two sub-networks have a significant influence on each other as they update themselves in an iterative manner. To learn a generator distribution over data $x$, the generator $G(z, \theta_1)$ maps the input noise variables $z$ to the expected data space $x$, while the discriminator $D(x, \theta_2)$ tries to learn the probability, ranging from $0$ to $1$, that the generated data are from the real domain. The weights of the discriminator are optimised to minimise the probability of fake samples being classified as belonging to the real dataset, and maximising the probability that real input images are classified that they are from the real dataset. As such, the discriminator is minimising
the function $D(G(z))$ while maximising $D(x)$. In contrast, the generator tries to defeat the discriminator by generating fake samples, which look similar to the real samples. Thus, the weights of the generator are optimised to increase the probability that the fake samples are classified as belonging to the real dataset.

As both generator and the discriminator train simultaneously, parameters are adjusted for generator to minimize $log(1 - D(G(x)))$
and for discriminator to minimize $logD(x)$, as if they are following the two-player min-max
game with value function $V(G, D)$.

The objective of a GAN can be formulated as,
\begin{align}
    \min_{G} \max_{D} V(D, G) = \mathbb{E}_{x \sim P_{\text{data}}(x)}[\log D(x)] \nonumber  \\ + \mathbb{E}_{z \sim P_{z}(z)}[\log(1 - D(G(x)))]~.
\end{align}
When the network converges, the generator can create more realistic samples. At this point, the discriminator becomes unable to differentiate between real and fake samples. This is because neither the discriminator nor the generator can further improve themselves.

\section{PCA for Scalability and Feature Learning}
\label{conventionalPCA}
In the proposed methodology, we address the scalability issue by transforming the input image data rather than breaking it into patches. We adopt PCA approach to extract and learn essential features of images that facilitates efficient learning. Initially, PCA normalises the input data, ensuring uniformity in feature scales. Subsequently, it decomposes images into principal components, which represent the most significant variations within the dataset. To understand how much of the information in the dataset can be captured by each principal component, we measure the explained variance ratio. Figure \ref{fig:scree-plot} shows the plot, where explained variance is 98\% with 40 principal components in PCA, indicating that these 40 components capture 98\% of the total variability  present in the original dataset. Therefore, we choose to extract 40 PCA features.

\textcolor{black}{Mathematically, given a dataset, $\mathbf{X} \in \mathbb{R}^{n \times d}$, where $n$ is the number of data points and $d$ is the dimensionality of the feature space, the covariance matrix of $X$ can defined as:
\begin{align}
\mathbf{C} = \frac{1}{n-1} \mathbf{X}^\top \mathbf{X}.
\end{align}
Performing Singular Value Decomposition (SVD) on $C$ yields:
\begin{align}
\mathbf{X} = \mathbf{U} \mathbf{\Sigma} \mathbf{V}^\top,
\end{align}
where $\mathbf{U} \in \mathbb{R}^{n \times n}$ represents the left singular vectors related to the principal components of the data, $\mathbf{\Sigma} \in \mathbb{R}^{n \times n}$ is a diagonal matrix of singular values which quantify the variance captured by each principal component, and $\mathbb{R}^{n \times n}$ contains the right singular vectors (eigenvectors of $C$).
The total variance in the data is proportional to the sum of all squared singular values can be expressed as:
\begin{align}
\text{Total Variance} = \sum_{i=1}^{r} \sigma_i^2.
\end{align}
To reduce dimensionality, we retain only the top $k$ principal components, where 
$k$ is chosen such that the cumulative explained variance ratio satisfies:
\begin{align}
\frac{\sum_{i=1}^{k} \sigma_i^2}{\sum_{i=1}^{r} \sigma_i^2} \geq \eta,
\end{align}
The reduced representation of the dataset for a given threshold $\eta$ is,
\begin{align}
\mathbf{X}_{\text{PCA}} = \mathbf{U}_k \mathbf{\Sigma}_k,
\end{align}
where $\mathbf{U}_k$ and $\mathbf{\Sigma}_k$ contain the first $k$ columns of $\mathbf{U}$ and $\mathbf{\Sigma}$, respectively.
After performing PCA, the features are distributed across 8 sub-generators and computation is performed parallelly. For each quantum sub-generator \( i \), the features are selected from specific indices based on the formula \( k = 4i \), where \( i \in \{0, 1, 2, \dots, 7\} \). For each \( i \)-th sub-generator, 5 features are selected from the PCA-transformed data by choosing the following indices as \{i, 39 - k, 38 - k, 37 - k, 36 - k\} These indices correspond to columns in \( X_{\text{PCA}} \), and the feature subset for each quantum sub-generator is given by:
\begin{align}
X_i = \Big\{ 
X_{\text{PCA}}[:, i], \, 
X_{\text{PCA}}[:, 39 - k], \, 
X_{\text{PCA}}[:, 38 - k], \notag \\
X_{\text{PCA}}[:, 37 - k], \, 
X_{\text{PCA}}[:, 36 - k] 
\Big\}
\end{align}
where \( X_i \) represents the subset of features allocated to the \( i \)-th quantum sub-generator. Before feeding the selected features into the quantum circuits, normalization is performed to ensure that all features lie within the same range. The minimum and maximum values of the PCA data are computed as:
\begin{align}
pca_{\text{min}} = \min(X_{\text{PCA}}), \quad pca_{\text{max}} = \max(X_{\text{PCA}})~.
\end{align}
These values are used to normalize the data across all quantum sub-generators, ensuring consistent input across all quantum circuits.}

\textcolor{black}{Although PCA can be computationally expensive for large datasets, the primary motivation for using PCA in our approach stems from the hardware constraints of current quantum computers, particularly the limited number of qubits available. In our method, PCA is not only used as a dimensionality reduction technique but also as a means to extract the most significant global features of the images. thereby improving the scalability of the quantum generator. By selecting the top 40 principal components, we retain over 98\% of the variance in the data while reducing the input dimensionality, which mitigates the computational cost associated with high-dimensional data.} These PCA features are then scaled between $0$ and $1$, enabling quantum sub-generators to operate across the entire input space. During the training, these scaled features are fed to the discriminator as ground truth to guide the generator to learn the distribution of the dataset. After training, the output are re-scaled to their original principal component range, and  inverse transformation is performed to reconstruct images with the original distribution. Although this approach overcomes the scalability issue and generate high-quality medical images, we investigate that, the distribution of PCA features can be unbalanced among multiple sub-generators, consisting of a variational quantum circuit (VQC) within the proposed approach. Conventionally, PCA features are distributed sequentially among the sub-generators, that is, the first few features are aligned to the first sub-generator, the next features are assigned to the next sub-generator and so on. This is illustrated in Figure \ref{fig:PCA-mode-collapse}(a). We develop our QIGL approach in such a way that only 5 qubits are required for image generation of a full-resolution of $4096$-pixel ($64\times 64$ pixel) image dataset. To distribute 40 PCA features, QIGL comprises of 8 sub-generators, each with a 5-qubit quantum circuit. Each sub-generator is assigned with 5 PCA features. As shown in Figure \ref{fig:PCA-mode-collapse}(a), the first 5 features are assigned to the first sub-generator ($G1$), the next 5 features are assigned to the second sub-generator ($G2$), while the last sub-generator ($G8$) is getting the last 5 PCA features. This leads to an imbalance in the distribution of features among the sub-generators because PCA features inherently exhibit a heavy concentration of explained variance in the initial features, with a notable drop-off in subsequent ones. As PCA features become entangled within the sub-generators to facilitate rich connections for image generation, if certain sub-generators lack useful information across all features, the overall utility of the generator diminishes. 

We address the imbalance distribution of PCA features by counteracting the uneven assignment of principal components to sub-generators as shown in Figure \ref{fig:PCA-mode-collapse}(b). Initially, we allocate the top principal component to the first sub-generator, followed by subsequent components assigned sequentially to each sub-generator until each holds one top principal component. Next, the remaining principal components are distributed in reverse order, assigning the last $n-1$ components to the first sub-generator, the second to last $n-1$ to the second sub-generator, and so forth until all features are allocated among the sub-generators as illustrated in Figure \ref{fig:PCA-mode-collapse}(c). This approach ensures a more balanced distribution, allowing all the sub-generators to contribute meaningfully during training without skewing the learning process.

\section{Mode Collapse in QIGL}
\label{mode-collapse}
Mode collapse in generative learning occurs when the generator fixates on producing a narrow range of outputs, often repeating similar patterns or generating limited types of samples. Figure \ref{fig:PCA-mode-collapse}(d) illustrates mode collapse, where many latent space $z$ can be mapped to only one real space $x$. However, it is crucial to maintain variety in image generation rather than only generating limited type of images from a particular class. QIGL addresses this issue by incorporating Wasserstein distance within the proposed framework. Wasserstein distance approximates Earth Mover’s Distance to make the generated image distribution equal to the real image distribution. This distance measures the minimum cost required to transform one distribution into another. The loss function is constructed using the  Kantorovich-Rubinstein duality as,
\begin{align}
W(P_r, P_{\theta}) = \sup_{\|f\|_{L} \leq 1} \mathbb{E}_{x \sim P_r} [f(x)] - \mathbb{E}_{x \sim P_{\theta}} [f(x)],
\end{align}
where $P_r$ represents the real data distribution and $P_\theta$ represents the generated data distribution parameterised by the generator network's parameters $\theta$. $\|f\|_{L}$ is the Lipschitz constant of the function $f$, The supremum (sup) is taken over all functions $f$ with $\|f\|_{L} \leq 1$. $\mathbb{E}_{x \sim P_r} [f(x)]$ represents the expected value of
$f(x)$) when $x$ is sampled from the real data distribution $P_r$, and $\mathbb{E}_{x \sim P_{\theta}} [f(x)]$ represents the expected value of $f(x)$ when $x$ is sampled from the generated data distribution, $P_{\theta}$, which is controlled by the generator network. This formulation allows for a more stable optimisation process compared to traditional GANs, as it encourages the generator to produce samples that are closer to the real data distribution.

\section{Classical GAN with PCA}
\label{classicalGAN-PCA}

\begin{figure*}
\begin{center}
\includegraphics[width=1\linewidth]{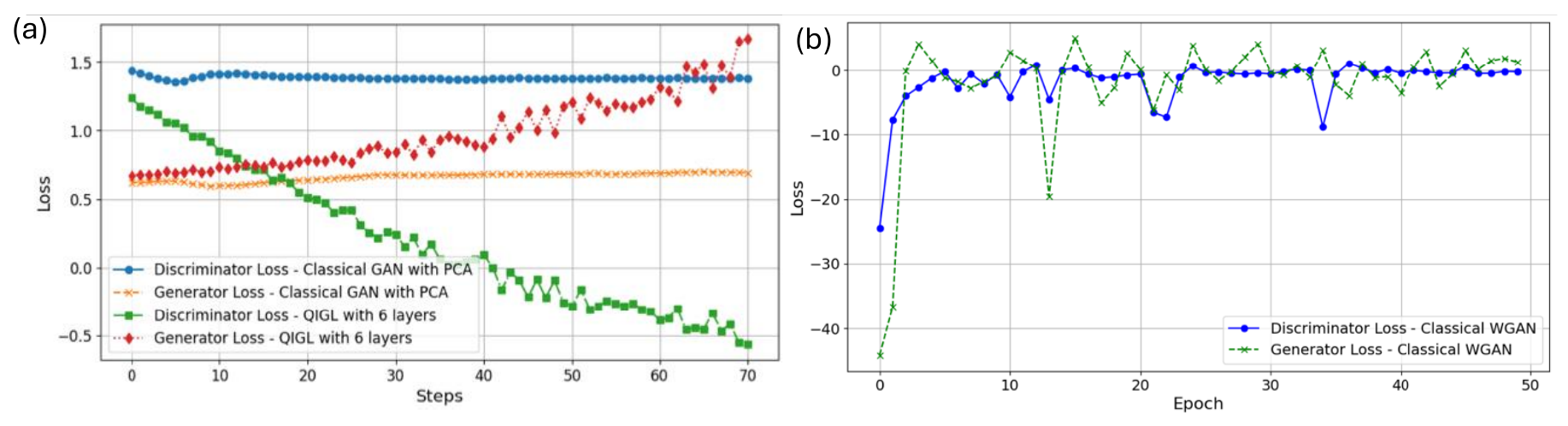}
\end{center}
\caption{\textcolor{black}{The generator and discriminator loss values during training of the models (a) QIGL model with 6 entangling layers, classical GAN model with PCA reduction, and (b) classical WGAN model. }}
\label{fig:lossvaluesplot}
\end{figure*}

{\color{black}{In the results and discussion section in the main manuscript, we compared the performance of our QIGL model to the existing quantum GAN models and classical WGAN model. In this work, we compare the proposed QIGL with traditional classical GAN without modifying the original architecture, i.e., without applying PCA, as classical GANs can directly handle high-dimesional data  without such limitations. However, in addition to this comparison, we extended our analysis to evaluate the performance of a classical GAN model when PCA is applied for feature reduction. We implemented a modified version of the classical generator for this analysis. It consists of a sequence of layers: an initial linear transformation from the latent dimension to 1024 dimensions, followed by another linear layer reducing it to PCA dimensions, and a final Tanh activation to ensure the output values are within the range of [-1, 1]. The model uses Leaky ReLU (Rectified Linear Unit) activations and optionally applies batch normalization to stabilize training. The total number of parameters in this classical generator is 144,424. We used the same classical discriminator that is used in the QIGL model.
The results are represented in Fig.\ref{fig:classicalGANwithPCA}. In this figure, we demonstrate the FID score values obtained during training the classical GAN model with PCA dimension of 40, and QIGL model with varying the number of entangling layers 4, 6, 8, and 10, respectively. The FID scores for the classical GAN model are still much higher compared to the values obtained by the QIGL model, which increase with increasing number of entangling layers (from 4 layers to 6 layers) and then remain in similar level (for 8 and 10 layers). These results demonstrate that the performance improvement is not solely attributable to PCA, thanks to expressiveness added by the entangling layers in the QIGL model and its hybrid framework. We also report the behavior of the discriminator and generator loss values of the models studied, i.e., classical WGAN model, QIGL model, and classical GAN model with PCA reduction in Figure \ref{fig:lossvaluesplot}.

{\color{black}{In this section, we also present the behavior of the discriminator and generator loss values of the models studied, i.e., classical WGAN model, QIGL model, and classical GAN model with PCA reduction, as seen in Fig.\ref{fig:lossvaluesplot}. These data demonstrate that QIGL converge faster than other models studied.}}

\begin{figure*}
\begin{center}
\includegraphics[width=0.75\linewidth]{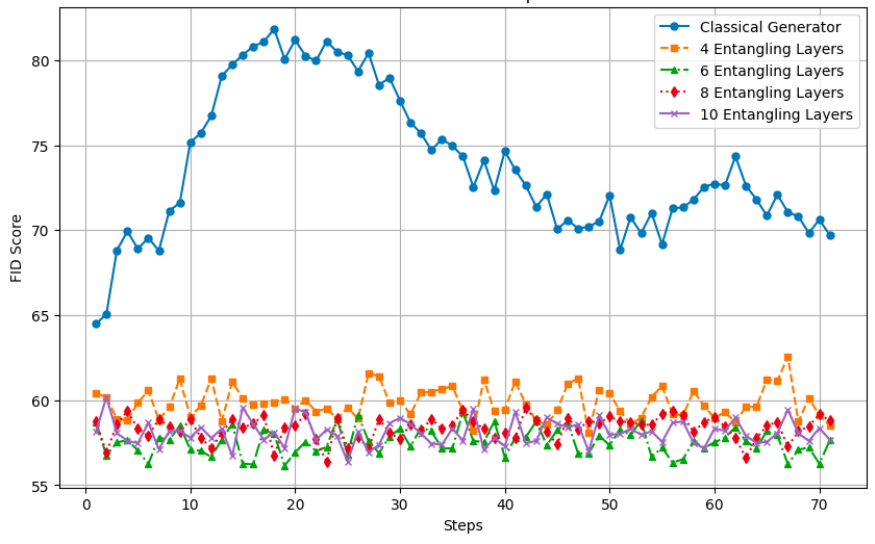}
\end{center}
\caption{\textcolor{black}{FID scores obtained during training of classical GAN model with PCA reduction, QIGL model with 4, 6, 8, and 10 entangling layers, respectively.}}
\label{fig:classicalGANwithPCA}
\end{figure*}}}

%These results demonstrate that the improvement in the FID score is thanks to expressiveness added by the entangling layers in the QIGL model and its hybrid framework.

%\section{Loss Values During Training}
%\label{lossvalues}
%{\color{red}{In this section, we present the behavior of the discriminator and generator loss values of the models studied, i.e., classical WGAN model, QIGL model, and classical GAN model with PCA reduction.}}

\end{document}